\documentclass[a4paper,12pt]{article}

\usepackage{amsmath,amssymb,epsfig,cite}

\newcommand{\beq}{\begin{equation}}
\newcommand{\eeq}{\end{equation}}
\newcommand{\ba}{\begin{array}}
\newcommand{\ea}{\end{array}}
\newcommand{\bea}{\begin{eqnarray}}
\newcommand{\eea}{\end{eqnarray}}
\newcommand{\bean}{\begin{eqnarray*}}
\newcommand{\eean}{\end{eqnarray*}}
\newcommand{\eref}[1]{(\ref{#1})}

\newcommand{\fref}[1]{Figure~\ref{#1}}
\newcommand{\nn}{\nonumber}
\newcommand{\tr}{\mathop{\rm Tr}}
\newcommand{\comment}[1]{}

\newcommand{\CM}{{\cal M}}
\newcommand{\CN}{{\cal N}}

\newcommand{\IP}{\mathbb{P}}
\newcommand{\IC}{\mathbb{C}}

\newcommand{\IZ}{\mathbb{Z}}
\newcommand{\f}{{\cal F}^{\flat}}

\newcommand{\setall}{\setcounter{equation}{0}
        \setcounter{theorem}{0}}

\begin{document}

\begin{flushright}
Imperial/TP/08/AH/10 \\
\end{flushright}
\vskip 0.25in

\renewcommand{\thefootnote}{\fnsymbol{footnote}}
\centerline{{\Huge M2-Branes and Quiver Chern-Simons:}}
\vspace{0.2cm}
\centerline{{\LARGE $\qquad \qquad$ A Taxonomic Study}}
~\\
\centerline{
Amihay Hanany${}^{1}$\footnote{\tt a.hanany@imperial.ac.uk, hanany@physics.technion.ac.il}
and
Yang-Hui He${}^{2}$\footnote{\tt hey@maths.ox.ac.uk}
}
~\\
~\\
{\hspace{-1in}
\scriptsize
\begin{tabular}{ll}
  ${}^1$ 
  &{\it Theoretical Physics Group, Blackett Laboratory, 
Imperial College, London SW7 2AZ, U.K. 
}\\
  ${}^2$
  & {\it Rudolf Peierls Centre for Theoretical Physics, Oxford University, 1 Keble Road, OX1 3NP, U.K.}\\
  & {\it Collegium Mertonense in Academia Oxoniensis, Oxford, OX1 4JD, U.K.}\\
  & {\it Mathematical Institute, Oxford University, 24-29 St.\ Giles', Oxford, OX1 3LB, U.K.}\\
\end{tabular}
}

\begin{abstract}
We initiate a systematic investigation of the space of 2+1 dimensional quiver gauge theories, emphasising a succinct ``forward algorithm''. Few ``order parametres'' are introduced such as the number of terms in the superpotential and the number of gauge groups. Starting with two terms in the superpotential, we find a generating function, with interesting geometric interpretation, which counts the number of inequivalent theories for a given number of gauge groups and fields. We demonstratively list these theories for some low numbers thereof. Furthermore, we show how these theories arise from M2-branes probing toric Calabi-Yau 4-folds by explicitly obtaining the toric data of the vacuum moduli space. By observing equivalences of the vacua between markedly different theories, we see a new emergence of ``toric duality''.
\end{abstract}

\setcounter{footnote}{0}
\renewcommand{\thefootnote}{\arabic{footnote}}

\newpage
\tableofcontents

\section{Introduction}\setall
Recent advances in studying the world-volume theory of M2-branes \cite{BL,gus} have allowed new perspectives on the AdS/CFT correspondence.  Indeed, the three-dimensional theory on the M2-branes is now understood as an ordinary gauge theory of Chern-Simons (CS) type \cite{VanRaamsdonk:2008ft,Aharony:2008ug}. The investigation of AdS$_4$/CFT$_3$ subsequently took flight, ever augmenting our arsenal of dual pairs of theories.

One strand of development has been the extension of the elaborate structure of the AdS$_5$/CFT$_4$ situation wherein the D3-brane world-volume gauge theory and the corresponding Calabi-Yau cone over Sasaki-Einstein 5-folds have been probed in great detail over the past decade. Of particular interest had been the cases where the Calabi-Yau threefold admits toric description. There, a rich tapestry, enabled by the abundant techniques of toric geometry, had been woven over such themes as toric duality \cite{Feng:2000mi,Feng:2002zw,Feng:2001xr}, dimer models and brane tilings \cite{Hanany:2005ve,Franco:2005rj,Feng:2005gw}.

Along this vein, the parallel story for M2-branes probing toric Calabi-Yau 4-fold singularities met with rapid progress \cite{Lambert:2008et,Hanany:2008cd,Hanany:2008fj,Franco:2008um,Benna:2008zy,Hosomichi:2008jb,Ueda:2008hx,Imamura:2008qs,Kim:2007ic,Lee:2007kv,Hanany:2008qc,Martelli:2008si}, addressing such issues as tiling or its 3-dimensional counter-part of crystal models, toric duality, as well as partition functions in the light of the plethystic programme \cite{pleth} etc. Thus inspired, especially by the fortitude to attack the general singularity \cite{Franco:2008um}, it is expedient for us to take a synthetic approach. The space of toric singularities of Calabi-Yau 4-folds is largely unchartered; nevertheless, we can gradually and systematically list, starting from the simplest imaginable, the possible $(2+1)$-dimensional quiver Chern-Simons theories. These theories are, of course, infinite in number; however, we will see how the number of nodes in the quiver and the number of terms in the superpotential can be used as order parametres to begin a classification. Even at low numbers, we will encounter many highly non-trivial theories. 
Furthermore, we will establish generating functions which counts the number of inequivalent theories at each level of our enumeration. Remarkably, we can geometrically interpret these functions. 

Thus, let us we embark on a journey through the complex {\it terra incognita} of new graphs and associated superpotentials, guided by our experience with the techniques of $(3+1)$-dimensional $\CN=1$ gauge theory, {\it mutatis mutandis}, and slowly probe the dual pairs of toric geometries and CS theories. A wealth of interesting properties will be encountered.

Our taxonomic study is organised as follows. We begin, in Section \ref{s:MCS}, with a brief review of the computation of the moduli space of quiver Chern-Simons theories in $(2+1)$-dimensions, culminating in an algorithmic flow-chart which generalises (and encompasses) the ``forward-algorithm'' of \cite{Feng:2000mi} to the present case. Next, in Section \ref{s:tax}, we begin our systematic investigation of what toric Chern-Simons quiver theories can exist. We start with two terms in the superpotential satisfying the toric condition and exhaustively present what possibilities could arise, using the number of nodes $G$ and the number of fields $E$ as order parametres. We then explicitly show how the moduli spaces of these theories are families of toric Calabi-Yau 4-folds, indexed by the Chern-Simons levels. We find a new manifestation of toric duality. In Section \ref{s:dimer}, we show how all these theories admit a dimer-model, or periodic-planar-tiling description, contrary to what one would naively expect. Finally, stepping back to have a bird's eye view, we find a generating function in Section \ref{s:count} which counts all the theories encountered. We conclude with prospect in Section \ref{s:conc}.

\section{Toric Geometry and CS Moduli Spaces}\label{s:MCS}\setall
In this section, we briefly outline the computation of moduli space of the theories of our interest: $(2+1)$-dimensional quiver Chern-Simons theories with $\CN=2$ supersymmetry. We point out that henceforth we are interested in the classical mesonic moduli space, which we will denote as $\CM$. 

The theories are with four supersymmetries and the gauge groups have no kinetic terms but instead have CS terms, moreover, they have product gauge groups together with bifundamental and adjoint matter. The computation of $\CM$ is a direct generalisation of the so-called ``{\bf forward algorithm}'' for $(3+1)$-dimensional $\CN=1$ gauge theories \cite{Feng:2000mi,Gray:2006jb}. Let the quiver CS theory have gauge group consisting of $G$ factors, and a total of $E$ fields, we then have the action, written in $\CN=2$ superspace notation:
\begin{equation}
L= -\int d^4 \theta\left( \sum\limits_{X_{ab}} X_{ab}^\dagger e^{-V_a} X_{ab} e^{V_b}
-i \sum\limits_{a=1}^G k_a \int\limits_0^1 dt V_a \bar{D}^{\alpha}(e^{t V_a} D_{\alpha} e^{-tV_a})
\right) + 
\int d^2 \theta W(X_{ab}) + c.c.
\end{equation}
where $a$ indexes the factors in the gauge group, $X_{ab}$ are the superfields accordingly charged, $V_a$ are the gauge multiplets, $D$ is the superspace derivative, $W$ is the superpotential and $k_a$ are the Chern-Simons levels which are integers; an overall trace is implicit since all the fields are matrix-valued. We take the following two constraints on the CS levels, the reader is referred to \cite{Hanany:2008cd,Hanany:2008fj} for details:
\begin{equation}\label{k-con}
\sum_{a=1}^{G} k_a = 0, \quad \gcd(\{k_a\}) = 1 \ .
\end{equation}

The classical mesonic moduli space $\CM$ is determined by the following equations
\begin{eqnarray}
\nn \partial_{X_{ab}} W &=& 0 \\
\nn \mu_a(X) := \sum\limits_{b=1}^G X_{ab} X_{ab}^\dagger - 
\sum\limits_{c=1}^G  X_{ca}^\dagger X_{ca} + [X_{aa}, X_{aa}^\dagger] &=& 
4k_a\sigma_a \\
\label{DF} \sigma_a X_{ab} - X_{ab} \sigma_b &=& 0 \ ,
\end{eqnarray}
where $\sigma_a$ is the scalar component of $V_a$. Indeed, this is in analogy to the F-term and D-term equations of $\CN=1$ gauge theories in $(3+1)$-dimensions, with the last equation being a new addition.

We are particularly interested in the case when $\CM$ is a toric variety where the forward algorithm conveniently uses the combinatorial power of lattice geometry and toric cones \cite{Douglas:1997de,Beasley:1999uz,Feng:2000mi}. This is the Abelian case where the gauge group is simply $U(1)^G$. Physically, $\CM$ is a toric Calabi-Yau 4-fold transverse to an M2-brane on whose world-volume lives the $(2+1)$-dimensional CS theory. For a stack of $N$ parallel, coincident M-branes, the moduli space is the $N$-th symmetrised product of (or more precisely the $N$-th Hilbert scheme of points on) the 4-fold \cite{Berenstein:2002ge,Aharony:2008ug,Hanany:2008cd,Hanany:2008fj}.

In this Abelian, toric case then, the third equation of \eref{DF} sets all $\sigma_a$ to a single field, say $\sigma$, on the coherent component of the moduli space. The second equation causes the D-terms to have FI-parametres $4k_a \sigma$.
The moduli space $\CM$ is a symplectic quotient of the space of solutions to the F-terms prescribed by the first equation modulo the gauge conditions prescribed by the D-terms. 
Because of the condition that all $k_a$ sum to 0 imposed in \eref{k-con} there is an overall $U(1)$ (corresponding to the center of mass motion of the M2-brane) which can be factored out. Furthermore, there is another $U(1)$ which can be factored. This is because the presence of CS couplings induces Fayet-Iliopoulos (FI) parameters on the space of D-terms. In a generic $(3+1)$-dimensional theory all these FI parameters are arbitrary, but here they are all aligned along a line which is parameterized by $\sigma$ and has a direction set by the CS integers. This picks out a very specific baryonic direction out of all possible directions which are present in $(3+1)$-dimensions. This direction becomes mesonic in $(2+1)$-dimensions and fibers over the Calabi-Yau 3-fold to give a total space as a Calabi-Yau 4-fold. Thus, in all, there is a net of $(G-2)$ D-terms. 

The space of solutions of the F-terms is itself a toric variety, of dimension $4+(G-2) = G+2$, this is the so-called ``{\bf Master space}'' $\f_{G+2}$, studied in detail in \cite{master}. Indeed, the $G-2$ D-terms are all the directions which remain baryonic in $(2+1)$-dimensions and give rise to the Master Space\footnote{There is a subtle point here. There are indeed $G-2$ baryonic directions for a given Lagrangian with $G$ gauge groups. This does not imply that all possible baryonic directions of the particular CY4 are given by these $G-2$ directions. It only provides a lower bound. There are at least $G-2$ such baryonic directions and a different formulation may give more than this number. Such a situation is evident from the study of models presented in \cite{Hanany:2008cd,Hanany:2008fj,Franco:2008um}.}. Since we are interested in the mesonic moduli space, we impose gauge invariance with respect to all of them.
In summary,
\begin{equation}\label{symp}
\CM_4 \simeq \f_{G+2} // U(1)^{G-2} \ ,
\end{equation}
where we have marked the complex dimensions explicitly as subscripts.
The $G-2$ FI-parametres are shown in \cite{Hanany:2008fj} to be in the integer kernel of the matrix
\begin{equation}\label{C}
C =\left(\begin{matrix}
1 & 1 & 1 & \ldots & 1 \\ k_1 & k_2 & k_3 & \ldots & k_G
\end{matrix}\right) \ .
\end{equation}

To proceed we now recall the essential features in the computation of the 3+1 dimensional mesonic moduli space of complex dimension 3.
Computationally, the charges of the fields are given by the so-called {\bf incidence matrix} of the quiver; this is a $G \times E$ matrix. Each column of this matrix representation of the quiver consists of two possible choices (1) a single pair $-1$ and $1$ denoting an arrow beginning and ending at the appropriate node and zero elsewhere, or (2) an entire column of zeros, denoting an adjoint field charged only under one single node. 

The incidence matrix is customarily denoted as $d$ and specifies the D-terms, i.e., the $U(1)$ toric actions. The crucial property of $d$ is that each of its columns sums to 0 (since each column corresponds to one arrow that necessarily has one head and one tail.) and that each of its rows also sums to 0 (this ultimately ensures that the moduli space be Calabi-Yau\footnote{Essentially, the reason is that one can translate the incidence matrix into the matrix of charges which define the toric variety by right multiplication of a perfect matching matrix as shown in the ensuing flow-chart. The Calabi-Yau condition, i.e., the vanishing of the first Chern class, is that the charges sum to zero and subsequently requires that the rows of the incidence matrix also sum to zero \cite{master}.}). 
We emphasise the ambiguity for an adjoint: we do not know under which precise node it is charged; from the point of view of the incidence matrix, there is no way to distinguish. We will see later how using information from the superpotential one may overcome this ambiguity.

Of course, one row of $d$ is redundant because of the summation rule, we delete it and customarily call the result $\Delta$. On the other hand, the F-terms can be solved in terms of a matrix $K$, whose dual cone we call $T$. From these we can extract so-called $U$ and $V$ matrices. We refer the reader to Section 2 of \cite{Feng:2000mi} for the details. Schematically we can summarise the procedure of the ``forward algorithm'', taken from {\it cit.~ibid.}, as follows:
\[
\begin{array}{ccccccc}
\mbox{Quiver} \rightarrow d_{G \times E}	& \rightarrow	&
	\Delta_{(G-1) \times E}	& & & & \\
	&	&\downarrow	&	&	&	&	\\
\mbox{F-Terms} \rightarrow K_{E \times (G+2)}	& \stackrel{V \cdot K^t =
	\Delta}{\rightarrow}
		& V_{(G-1)\times(G+2)}	 & & & & \\
\downarrow	&	& \downarrow	& & & & \\
T_{(G+2)\times c } = {\rm Dual}(K)	& \stackrel{U \cdot T^t = {\rm
	Id}}{\rightarrow} & U_{(G+2)\times c} & \rightarrow & VU & &\\
\downarrow	&	&	&	& \downarrow	& & \\
Q_{(c-G-2)\times c} = [{\rm Ker}(T)]^t	&	& \longrightarrow	& & (Q_t)_{(c-3) \times c} =
\left( \begin{array}{c}
(VU)_{(G-1)\times c} \\
Q_{(c-G-2) \times c} \\ 
\end{array} \right) & 
\end{array}
\]
We have marked the dimensions of each matrix for clarity. Note that $c$ is the number of perfect matching in the dimer model \cite{Hanany:2005ve} description of the theory to which we shall later turn. The key point is that we can combine the matter content (specified by the incidence matrix) and the superpotential (specified by the $K$-matrix) into a single charge matrix $Q_t$; its kernel, $G_t$, of dimension $3 \times c$, encodes the toric diagram of the Calabi-Yau threefold moduli space. Refreshed with this recollection let us make one more step before getting back to the main case of interest.

When dealing with 2+1 dimensional Chern-Simons theory the moduli space is a toric 4-fold. The above procedure should be modified to include the CS-levels by incorporating the matrix $C$. First, let us recast the above flow-chart for a 3+1 dimensional theory in a more succinct manifestation, dispensing of the need of the $V$ and $U$ matrices and introducing the perfect-matching matrix $P_{E \times c} = K \cdot T$ and the trivial matrix $(1,1,1, \ldots, 1)_{1 \times G}$, which by abuse of notation for now, we also call $C$:
\[
\begin{array}{lllllll}
\begin{array}{l}
\mbox{INPUT 1:} \\
\mbox{   Quiver}
\end{array}
& \rightarrow & d_{G \times E}	& \rightarrow	&
	(Q_D)_{(G-1) \times c} = 
	\ker(C)_{(G-1)\times G} \cdot \tilde{Q}_{G \times c} \ , \quad
(d_{G \times E} := \tilde{Q}_{G\times c}\cdot (P^T)_{c \times E})\\
&&\\[-0.5cm]
&&& \nearrow &&&\\
\begin{array}{l}
\mbox{INPUT 2:} \\
\mbox{   Superpotential} 
\end{array}
& \rightarrow & P_{E \times c}	& \rightarrow
		& (Q_F)_{(c-G-2)\times c} = [\ker P]^t\\
&&&&~~~~~~\downarrow\\
&&&& (Q_t)_{(c-3) \times c} =
\left( \begin{array}{c}
(Q_D)_{(G-1)\times c} \\
(Q_F)_{(c-G-2) \times c} 
\end{array} \right) \rightarrow
\begin{array}{l}
\mbox{OUTPUT: } \\
(G_t)_{3 \times c} = [{\rm Ker}(Q_t)]^t\\
\end{array}
\end{array}
\]
\comment{
Q has c x (c-G-2) where c is the number of pm's.
Qt has c x (c-4). we add G-2 D term conditions.
C has size 2 x G
ker C has G x (G-2)
So generically we can not multiply ker C and Qt - sizes do not match.
}
Note that we have judiciously called the charges coming from the D-terms $Q_D$ and those from the F-terms, $Q_F$. Moreover, we have even foregone the need for the matrix $K$, perhaps so deeply engrained from our incipient days, and worked entirely in terms of the perfect-matching matrix $P$; there is indeed a conducive algorithm for extracting $P$ directly from the superpotential (cf.~\cite{Franco:2008um}). For the reader's convenience, we summarise this in Appendix \ref{a:P}.

Thus written, the generalisation to the case of present interest, viz. 2+1 dimensional CS theories, is most straight-forward. We only need to modify the $C$ matrix from the row of ones to our current $C_{2 \times G}$ matrix defined in \eref{C} which now includes the CS levels, thereby changing the dimension of $Q_D$ by 1, and subsequently causes $G_t$ to have 4 rows:
\begin{equation}\label{Gt}
\begin{array}{lllllll}
\fbox{
\mbox{
\begin{tabular}{l}
INPUT 1: \\
~~Quiver
\end{tabular}
}}
& \rightarrow & d_{G \times E}	& \rightarrow	&
	(Q_D)_{(G-2) \times c} = 
	\ker(C)_{(G-2)\times G} \cdot \tilde{Q}_{G \times c} \ ; 
&&\\[-0.3cm]
\vspace{-0.5cm}&&& \nearrow & \qquad \mbox{ with } d_{G \times E} := \tilde{Q}_{G\times c}\cdot (P^T)_{c \times E} &&\\
\fbox{
\mbox{
\begin{tabular}{l}
INPUT 2: \\
~~CS Levels
\end{tabular}
}}
& \rightarrow & C_{2 \times G}	&&&&\\[-0.3cm]
&&& \nearrow &&&\\
\fbox{\mbox{
\begin{tabular}{l}
INPUT 3: \\
~~Superpotential
\end{tabular}
}}
& \rightarrow & P_{E \times c}	& \rightarrow
		& (Q_F)_{(c-G-2)\times c} = [\ker P]^t \ ; \\
&&&&~~~~~~\downarrow\\
&&&& 
\hspace{-1in}
(Q_t)_{(c-4) \times c} =
\left( \begin{array}{c}
(Q_D)_{(G-2)\times c} \\
(Q_F)_{(c-G-2) \times c} 
\end{array} \right) \rightarrow
\fbox{\mbox{
\begin{tabular}{l}
OUTPUT:  \\
~~$(G_t)_{4 \times c} = [{\rm Ker}(Q_t)]^t$\\
\end{tabular}
}}
\end{array}
\end{equation}
The matrix $G_t$ represents the desired toric diagram: it should have columns of length 4, signifying a 4-fold; moreover, these 4-vectors should be co-spatial, i.e., they all live in a 3-dimensional hypersurface, this is required by the Calabi-Yau condition. There could be repetitions out of the $c$ columns, this is the multiplicity, or perfect matchings, discussed in \cite{Feng:2000mi,Feng:2001xr,Feng:2001bn,Hanany:2005ve,Sarkar:2000iz,Agarwal:2008yb,Muto:2002pk}.

\section{A Taxonomic Study}\label{s:tax}\setall
Having outlined the computational procedure, let us now proceed with a systematic study of examples. Ideally, one would wish for a classification of all possible theories, their quiver diagrams, interactions and subsequent geometries. This is a daunting task of organising an infinite number of models. Nevertheless, we can proceed cautiously and modestly: let us start with theories with a single pair of black-white nodes in the dimer picture. These correspond to cases where the superpotential vanishes for a single brane; this is because the so-called {\bf toric condition} \cite{Feng:2002zw} requires that each field appears exactly twice with opposite signs. In terms of the dimer model \cite{Hanany:2005ve}, this condition is what gives rise to the bi-partite nature of the tiling.
Of course, for multiple branes, because the fields become matrix-valued and do not commute necessarily, the superpotential may no longer vanish. This is familiar to us. For example, in the 3+1 dimensional gauge theory for the conifold 3-fold singularity there is a quartic superpotential which vanishes for a single brane. We will encounter this situation in detail below.

Hence, with a single M2 brane, and with only two terms therein, the superpotential actually vanishes, subsequently, the master space is freely generated by the $E$ fields and is simply $\IC^{E}$. Thus, the matter content completely specifies the 4-fold singularity. The symplectic quotient \eref{symp} and the traditional approach both become relatively simple in this case. In fact, in \eref{Gt}, $c=E$ and the perfect matching matrix $P$ is just the identity matrix. The $Q_F$ matrix for the F-terms are not present and we have that:
\begin{equation}\label{symp2}
\CM \simeq \IC^E // U(1)^{G-2} \sim G_t = 
\ker(\ker(C)_{(G-2) \times G}\cdot d_{G \times E}) \ ,
\end{equation}
where $\sim$ means $\CM$ has the toric diagram given by $G_t$ of dimensions $4 \times E$.

Indeed, from \eref{symp}, we see that $E = G+2$; let us hence use $E$ as a single order parametre and proceed gradually. We initiate with the case of two nodes, i.e., $(G,E) = (2,4)$. For each incremental value of $E$, we classify all $d$-matrices satisfying the constraints which define the incidence matrix. Indeed, in \eref{symp2}, $\ker(C) \times d$ has dimensions $(G-2) \times (G+2)$ and its kernel $G_t$ will have columns of length 4, as is required for a toric diagram of a 4-fold.

In summary, our classification scheme for 2 terms in the superpotential proceeds as follows:
\begin{enumerate}
\item Fix $G$, the number of nodes in the quiver. This also fixes the number of arrows as $E = G+2$; 
\item Find all $G \times (G+2)$ matrices which are incidence matrices, i.e., (a) each column is one of only zeros (adjoint) or consists of a single pair of $-1$ and $1$ and zero otherwise (bi-fundamental), (b) each row sums to 0 (Calabi-Yau condition);
\item Identify all loops (gauge invariant operators) in the quiver drawn from each incidence matrix and construct possible 2-term superpotentials satisfying (a) the toric condition, i.e., each bi-fundamental field occurs exactly twice and with opposite sign and (b) for a single brane when all the fields reduce from matrix-valued to complex numbers, the 2 terms conspire to cancel.
\end{enumerate}

How do we attack step 2, the most computationally intense one?
Luckily, the algorithms for such a matrix-partitioning problem were implemented in \cite{Anderson:2008uw} and we can thus happily proceed with presenting the solutions.  We adhere to the standard notation that 
\begin{quote}
{\it For bi-fundamentals $X^{i}_{ab}$ denotes the $i$-th arrow from node $a$ to $b$ and similarly that $\phi^i_a$ denotes the $i$-th adjoint on node $a$ (when there is only a single arrow the $i$-index is dropped).}
\end{quote}
Already, there are many highly non-trivial theories. 

\subsection{Four Fields in the Quiver}
Here, there are 2 nodes and we find only two non-isomorphic solutions.
These are presented in \fref{f:E=4} in the Appendix. There are two possibilities. In Model (1) we see that this is a non-chiral theory with 4 bi-fundamentals. We recognise the quiver as that of the conifold quiver. Indeed, the full superpotential for arbitrary $N$ of that theory also has 2 terms: calling the bi-fundamentals $X_{12}^i$ and $X_{21}^i$, ($i=1,2$ and $U(N)$ matrices) we have that $W = \tr(X^1_{12} X^1_{21} X_{12}^2 X^2_{21} - X^1_{12} X^2_{21} X_{12}^2 X^1_{21})$, which indeed satisfies the toric condition and vanishes for $N=1$. The fact that we have CS constraints, of course, modifies the moduli space from the 3-dimensional conifold to a CY 4-fold.

In Model (2), there are 2 adjoint fields and 2 bi-fundamentals.
Where shall we place the 2 non-bi-fundamental fields in Model (2)? We can be guided by promoting to $N>1$ number of branes. There could be two possible placements: both as adjoints on 1 node or one on each node. Calling the two bi-fundamentals $X_{12}$ and $X_{21}$, for both adjoints on the same node, say node 2 without loss of generality, we have 2 gauge invariant terms: $\tr(X_{12}\phi_2^1\phi_2^2X_{21})$ and $\tr(X_{12}\phi_2^2\phi_2^1X_{21})$. On the other hand, for one adjoint on each node we have only one possible invariant: $\tr(X_{12}\phi_2X_{21}\phi_1)$; this is because we need to be careful in matrix composition when we go about the loops in the quiver to construct the gauge invariants. Therefore, Model (2) has a natural candidate for a 2-term superpotential satisfying the toric condition and having two terms, namely $W = \tr(X_{12}[\phi_2^1,\phi_2^2]X_{21})$.

\begin{figure}[ht]\begin{center}
\epsfxsize = 10cm\epsfbox{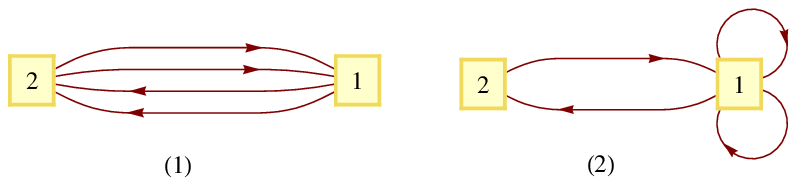}
$W_{(1)} = \tr(X^1_{12} X^1_{21} X_{12}^2 X^2_{21} - X^1_{12} X^2_{21} X_{12}^2 X^1_{21})$; $\qquad$
$W_{(2)} = \tr(X_{12}[\phi_2^1,\phi_2^2]X_{21})$
\caption{{\sf The quivers with 4 fields and 2 nodes. There are 2 solutions and the 2-term superpotentials are also given. The moduli space in both cases is just the trivial CY 4-fold $\IC^4$.
}}
\label{f:E=4full}
\end{center}\end{figure}

In summary, the two full models are drawn in \fref{f:E=4full}.
Of course, the actual moduli space $\CM$ for both models are easily determined using \eref{symp2}. Here $G-2 = 0$ and there is no symplectic action. Thus the moduli space is simply the master space, viz., $\IC^4$. The toric diagram is merely 4 corners of a tetrahedron.

We are assured by the fact that these two models have appeared in the literature, Model (1), in \cite{Aharony:2008ug} and Model (2), in \cite{Hanany:2008fj,Franco:2008um}, in each case from different lines of reasoning. Here we have arrived at these from yet another viewpoint -- a systematic scan of theories.

It is perhaps important to emphasize that Model (2) is not considered to be a consistent model in 3+1 dimensions even though it does admit a two dimensional tiling \cite{Hanany:2008fj} (here consistent is taken to mean that it leads to a 3+1 dimensional SCFT under the RG flow to the IR, with a known AdS dual). The essential feature of Model (2) is node 2 which has 1 flavor (or taking the SQCD conventions of the number of flavors $N_{f}=N_{c}$ equal to the number of colors). Henceforth we shall call such nodes as ``{\bf one-flavored nodes}''. Indeed it is a standard lore that such theories develop a scale and confine in 3+1 dimensions, thus not leading to SCFT. On the other hand, such theories in 2+1 dimensions do not necessarily develop a scale \cite{Aharony:1997bx} and can perfectly lead to a non-trivial SCFT under the RG flow. Once we allow for theories with one-flavored nodes, a whole space of opportunities reveals itself and a zoo of new SCFT's in 2+1 dimensions becomes available. We encounter more examples of this type in the following sections.

\subsection{Five Fields in the Quiver}
Here, there are 3 nodes and we find 5 distinct solutions. These are shown in \fref{f:E=5} in the Appendix. Model (1) has 5 bi-fundamentals. Models (2) and (3) have 4 bi-fundamentals and 1 adjoint field. Note that Model (2) has a disconnected node labelled 2. Model (4) has 3 bi-fundamentals and 2 adjoint fields and finally Model (5), also with a disconnected node, has only 2 bi-fundamentals and 3 adjoint fields.
Let us ignore models (2) and (4) since they inevitably have reducible moduli spaces due to disconnected nodes. 

For Model (1), using the standard notation, we clearly have the 2-term superpotential: $W = \tr(X_{21}X_{13}^1X_{31}X_{13}^2X_{32} - X_{21}X_{13}^2X_{31}X_{13}^1X_{32})$, which satisfies the toric condition and vanishes for $N=1$ when all fields are simply complex numbers.

For Model (3), clearly there are two invariants $X_{21}X_{12}$ and $X_{13}X_{31}$. Where shall we place the missing adjoint $\phi$? It is easy to see that placing it on node 2 gives two invariants once we promote to $N>1$ branes, i.e., to matrix-valued fields: $\tr(X_{21}\phi_1 X_{13} X_{31} X_{12})$ and $\tr(X_{21}X_{13} X_{31} \phi_1 X_{12})$. This indeed gives a two-term superpotential satisfying the toric condition of each field appearing exactly twice with opposite signs: $W = \tr(X_{21}\phi_1 X_{13} X_{31} X_{12} - X_{21}X_{13} X_{31} \phi_1 X_{12})$; moreover, $W$ indeed vanishes at $N=1$ when all field are merely complex numbers.

\begin{figure}[t]\begin{center}
\epsfxsize = 14cm\epsfbox{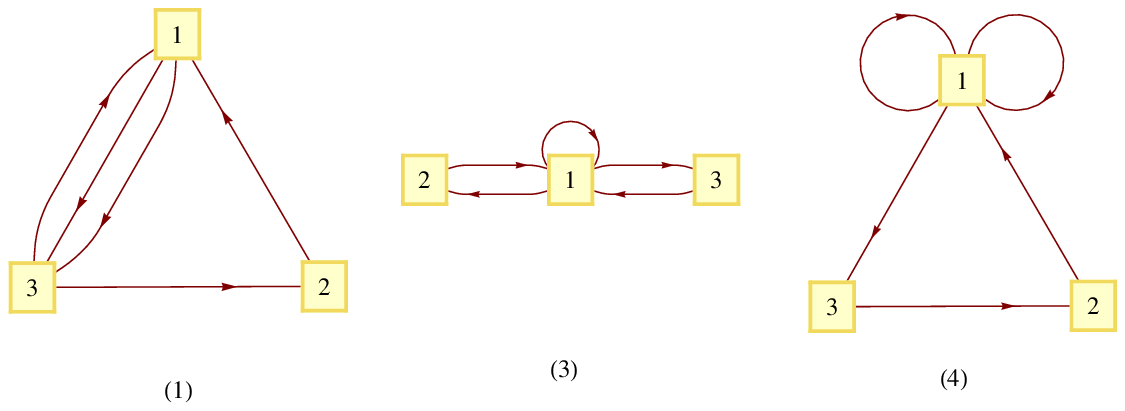}
$\begin{array}{l}
W_{(1)} = \tr(X_{21}X_{13}^1X_{31}X_{13}^2X_{32} - X_{21}X_{13}^2X_{31}X_{13}^1X_{32}) \ ; \\
W_{(3)} = \tr(X_{21}\phi_1 X_{13} X_{31} X_{12} - X_{21}X_{13} X_{31} \phi_1 X_{12}) \ ; \\
W_{(4)} = \tr(X_{21}[\phi^1_1, \phi^2_1]X_{13}X_{32}) . \ \\
\end{array}$
\caption{{\sf The quivers with 5 fields and 3 nodes. There are 3 good models and we call them (1), (3) and (4). We also write the corresponding 2-term superpotentials.
}}
\label{f:E=5full}
\end{center}\end{figure}

Finally, for Model (4), there are 2 ways of placing the 2 adjoints: both on the same node, say node 1, or one each on 2 of the nodes. The latter possibility is ruled out since it would be impossible to have 2 terms and satisfying the toric condition. However, putting both $\phi_1^{i=1,2}$ on node 1, gives us a superpotential satisfying all requisites: $W = \tr(X_{21}[\phi^1_1, \phi^2_1]X_{13}X_{32})$.
We summarise all these good models in \fref{f:E=5full}. 

It is important to note that none of these 3 models appeared previously in the literature. Indeed, our systematic study revealed the existence of these 3 prototypical models. We will proceed later by analyzing their properties and further introduce them from yet another taxonomic viewpoint.
It is further important to notice that all these models have a one-flavored node and therefore do not correspond to consistent models in 3+1 dimensions even though they all admit a brane tiling. Nevertheless they present a rich structure of non-trivial SCFT's in 2+1 dimensions with a highly intricate spectrum of scaling dimensions.

\subsection{Six Fields in the Quiver}
\begin{figure}[ht!!!]\begin{center}
\epsfxsize = 14cm\epsfbox{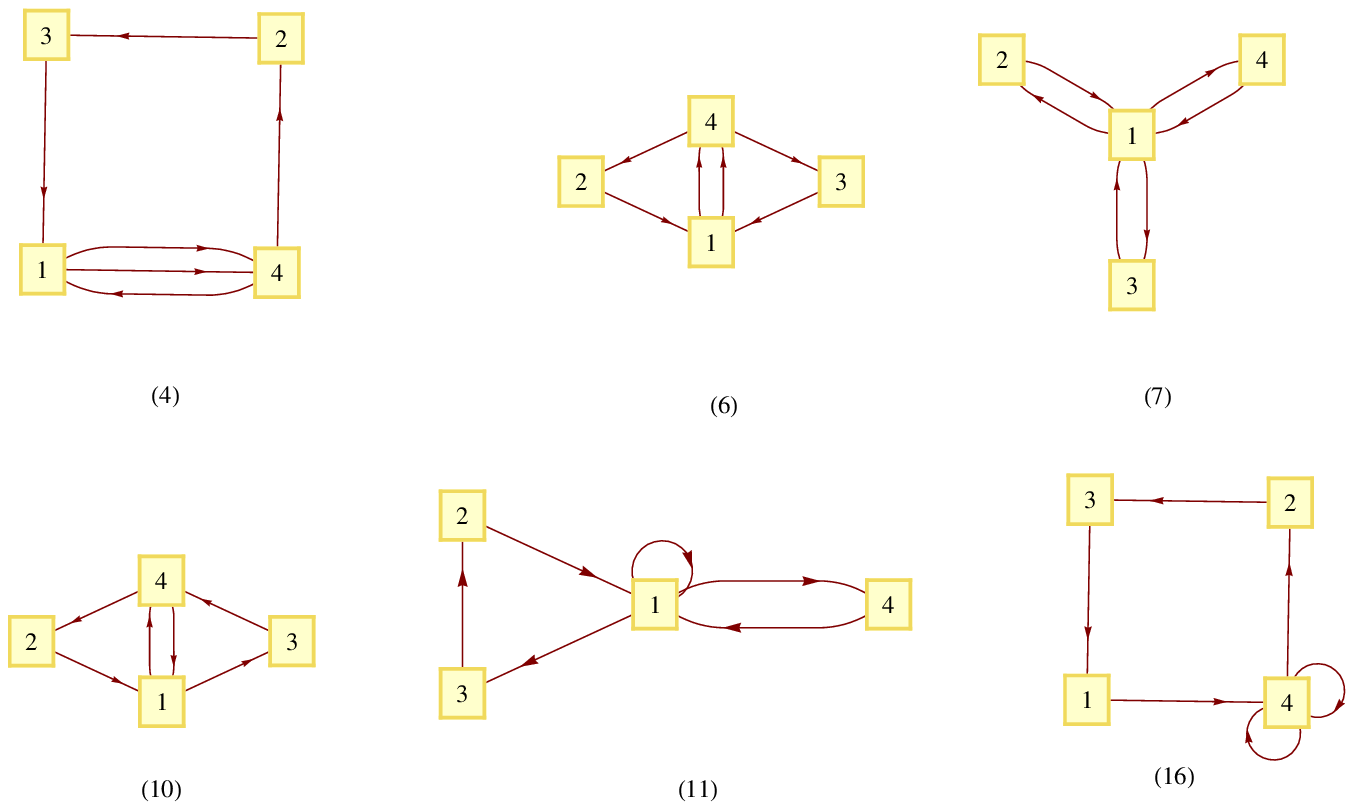}
$
\begin{array}{l}
W_{(4)} = \tr(X_{31}X_{14}^1X_{41}X_{14}^2X_{42}X_{23} - X_{31}X_{14}^2X_{41}X_{14}^1X_{42}X_{23}) \ ; \\
W_{(6)} = \tr(X_{42}X_{21} (X_{14}^1 X_{43}X_{31}X_{14}^2 - X_{14}^2 X_{43}X_{31}X_{14}^1) ) \ ; \\
W_{(7)} = \tr(X_{12}X_{21} (X_{14}X_{41}X_{13}X_{31} - X_{13}X_{31}X_{14}X_{41}) ) \ ; \\
W_{(10)} = \tr(X_{42}X_{21} X_{14} X_{41}X_{13}X_{34} - X_{42}X_{21}X_{13}X_{34}X_{41}X_{14}) \ ; \\
W_{(11)} = \tr(X_{32}X_{21}\phi_1X_{14}X_{41}X_{13} - X_{32}X_{21}X_{14}X_{41}\phi_1 X_{13} ) \ ; \\
W_{(16)} = \tr(X_{42}X_{23}X_{31}X_{14}[\phi_4^1, \phi_4^2])
\end{array}
$
\caption{{\sf The quivers with 6 fields and 4 nodes. There are 6 models. We also present the superpotential as well as the adjoint-fields where necessary.
}}
\label{f:E=6full}
\end{center}\end{figure}

Here, there are 4 nodes and there is a total of 18 distinct solutions. These are shown in \fref{f:E=6} in the Appendix. Models (1) to (10) have 6 bi-fundamentals. Note that Models (1), (2), (5) and (9) all have disconnected nodes. Model (8), though apparently acceptable, has a superpotential in terms of its bi-fundamentals which looks like $ \tr(X_{31}X_{14}X_{42}X_{24}X_{41}X_{13} - X_{42}X_{24}X_{41}X_{13}X_{31}X_{14} )$. This vanishes by cyclicity of the trace and hence it is really just a theory without superpotential at all. Models (11) and (12) have 5 bi-fundamentals and 1 adjoint field. Models (13) to (16) have 4 bi-fundamentals and 2 adjoint fields. Note that Models (13) and (14) both have disconnected nodes. Model (17) has 3 bi-fundamentals and 3 adjoint fields, as well as a disconnected node. Finally, Model (18) has 2 bi-fundamentals, 4 adjoint fields and 2 disconnected nodes.
Once again, we select the diagrams without detached nodes, insert the appropriate adjoint fields, and also write down the possible 2-term superpotentials. We find a total of 6 possible models. These are presented in \fref{f:E=6full}.

Happily, we have again recovered and extended some of the known models in the literature. Model (6) was proposed in Figure 3 of \cite{Franco:2008um} for the $Q^{1,1,1}$ geometry while Model (10) was proposed in Figure 7 of \cite{Franco:2008um} for the so-called $D_3$ theory. The other models appear for the first time in the literature. All models have a one-flavored node and so none correspond to consistent models in 3+1 dimensions even though they all admit a brane tiling description.

\begin{figure}[ht!!!]\begin{center}
\epsfxsize = 14cm\epsfbox{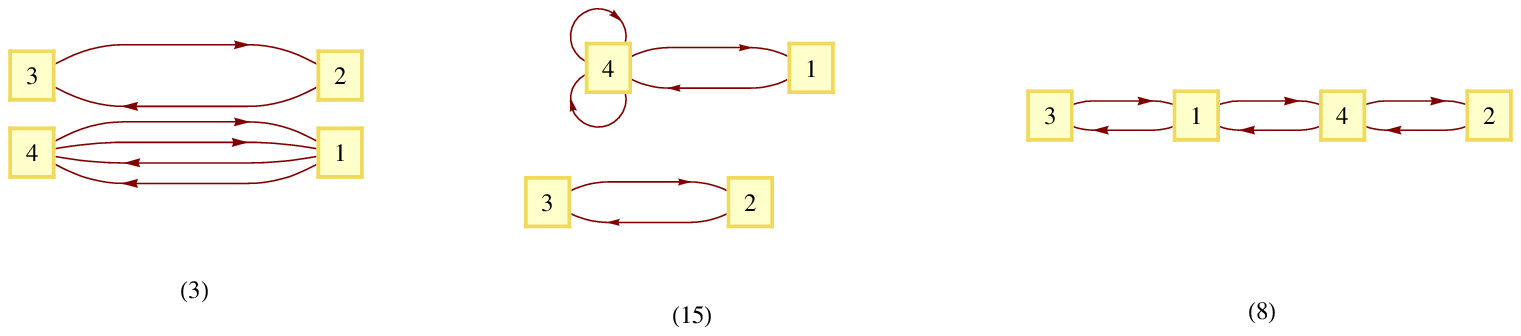}
$
\begin{array}{l}
W_{(3)} = \tr(X_{23}X_{32})\tr(X^1_{14} X^1_{41} X_{14}^2 X^2_{41} - X^1_{14} X^2_{41} X_{14}^2 X^1_{41})) \ ; \\
W_{(15)} = \tr(X_{23}X_{32})\tr(X_{14}[\phi_4^1,\phi_4^2]X_{41}) \ ; \\
W_{(8)} = \tr(X_{31}X_{14}X_{42}X_{24}X_{41}X_{13} - X_{42}X_{24}X_{41}X_{13}X_{31}X_{14} ) = 0 \ ; \\
\end{array}
$
\caption{{\sf The two disconnected quivers with 6 fields and 4 nodes, together with their multi-trace superpotentials. Also Model (8) has completely vanishing
superpotential.
}}
\label{f:E=6disconnect}
\end{center}\end{figure}

For completeness we also include the two disconnected quivers, Models (3) and (15). These engender multi-trace superpotentials because of the quivers factorise. We present these in \fref{f:E=6disconnect}. We see that they are essentially products of Models (1) and (2) of the 4-edged case, together with a 2-edged bi-fundamental non-chiral quiver. Also, we include Model (8), which has a completely vanishing superpotential if it were to be 2-termed. We will exclude these models from discussion in the following.

\subsection{Toric Diagrams: Illustrious Examples}
Let us take, for concreteness, Model (1) of the $E=5$ quivers, given in \fref{f:E=5full}. There are 3 nodes and hence 2 independent CS levels, $k_1$ and $k_2$.
The toric diagram, according to \eref{Gt}, is given by
\begin{eqnarray}
\nn
&G_t^{(1)}& = \ker(\ker
\left(\begin{matrix}
1 & 1 & 1 \\ k_1 & k_2 & -k_1-k_2 
\end{matrix}\right)
\cdot
{\scriptsize \left(\begin{matrix}
-1 & -1 & 0 & 1 & 1 \\
 0 & 0 & 1 & -1 & 0 \\
 1 & 1 & -1 & 0 & -1
\end{matrix}\right) }) = 
\ker
\left(
 -k_2 , -k_2 , -k_1 , k_1+k_2 , k_2
\right) \\
&=&
\mbox{gen}{\tiny
\left(
\begin{array}{lllll}
 k_2 & 0 & 0 & 0 & k_2 \\
 k_1+k_2 & 0 & 0 & k_2 & 0 \\
 -k_1 & 0 & k_2 & 0 & 0 \\
 -k_2 & k_2 & 0 & 0 & 0
\end{array}
\right)} =
{\tiny
\left(
\begin{array}{lllll}
 1 & 0 & 0 & 0 & 1 \\
 k_1+k_2 & 0 & 0 & k_2 & 0 \\
 -k_1 & 0 & k_2 & 0 & 0 \\
 -1 & 1 & 0 & 0 & 0
\end{array}
\right)}
\ ; \ \ \gcd(k_1,k_2,k_1+k_2) = 1 \ .
\end{eqnarray}
In the penultimate step, we have been mindful of the fact we need to find the integer kernel of the charge matrix and not merely the nullspace and  hence we wrote gen(~) therein to denote that we should reduce to a basis over the integers for whichever choice of $k_1$ and $k_2$. This means that each row of $G_t$ must have GCD being 1. Guaranteed by the condition $\gcd(k_1,k_2,-k_1-k_2) = 1$, rows 2 and 3 are acceptable but rows 1 and 4 need to divide out the common factor of $k_2$; this gives the last step.

Indeed, each column of $G_t$ is of length 4, signifying that the resulting moduli space corresponding to this theory is a toric 4-fold. Furthermore, we see that the vector $(1,1,1,1)$ is perpendicular to every pair-wise linear combination between the 5 column vectors, this means that the columns are actually co-spatial, i.e., live on a dimension 3 hypersurface in $\IZ^4$; this, of course, guarantees that our 4-fold is in fact Calabi-Yau. Thus re-assured, we can, without loss of generality, delete any row of $G_t$ (call it $G_t'$) and represent the toric 4-fold by an integer polytope in 3-dimensions. Thus, we can write, for the toric diagram,
\begin{equation}
G_t' = \mbox{gen}{\scriptsize
\left(
\begin{array}{lllll}
 1 & 0 & 0 & 0 & 1 \\
 -k_1 & 0 & k_2 & 0 & 0 \\
 -1 & 1 & 0 & 0 & 0
\end{array}
\right)} \ , \qquad \mbox{ with } \gcd(k_1,k_2,k_1+k_2) = 1 \ ,
\end{equation}
where for convenience we have deleted the second row.

\begin{figure}[ht]\begin{center}
\epsfxsize = 12cm\epsfbox{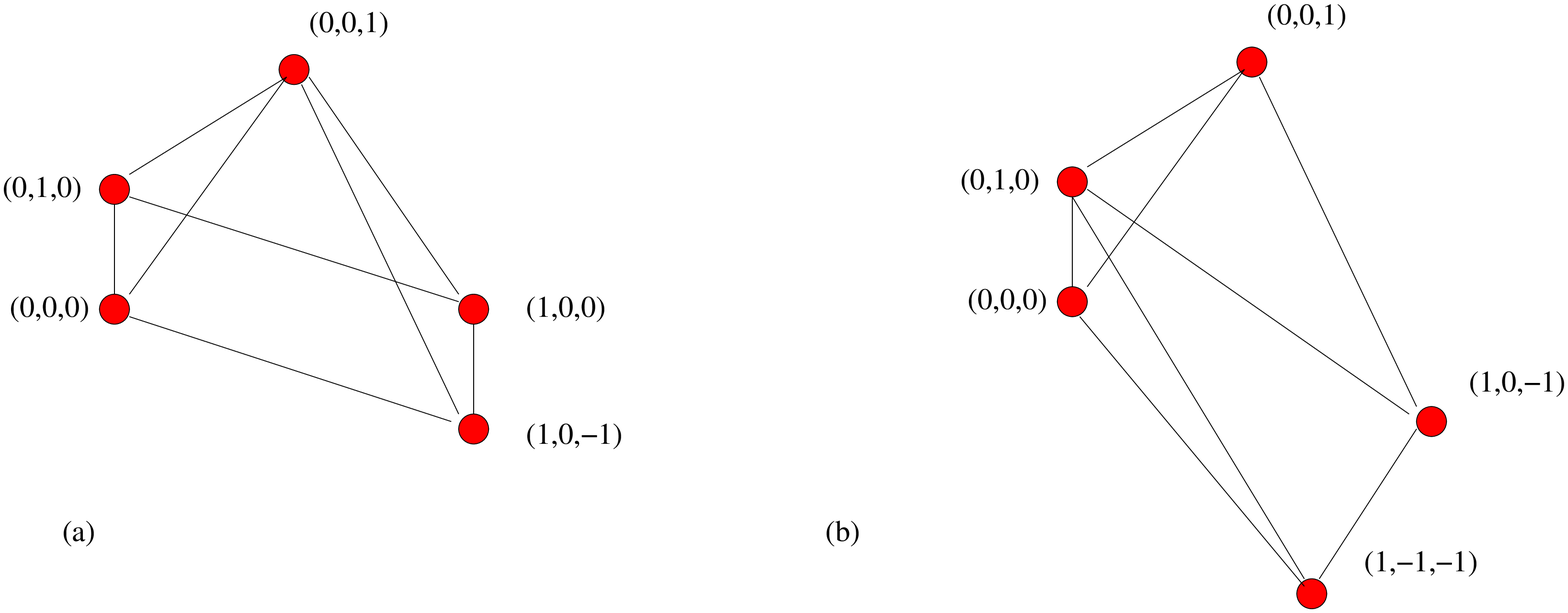} 
\caption{{\sf The two moduli spaces, drawn with explicit toric diagrams, for different choices of Chern-Simons levels (a) $(k_1,k_2)=(0,1)$ and (b) $(k_1,k_2)=(1,1)$ for Model (1) of the 5-edge, 3-noded theory with 2-term superpotential.
}}
\label{f:toricE=5}
\end{center}\end{figure}

Now, depending on the choice of $k_1$ and $k_2$ obeying the coprimarity condition, we have an infinite family of toric CY4s. As two illustrious examples we have that
\begin{equation}
G_t'(k_1 = 0, k_2 = 1) = 
\left({\scriptsize
\begin{array}{lllll}
 1 & 0 & 0 & 0 & 1 \\
 0 & 0 & 1 & 0 & 0 \\
 -1 & 1 & 0 & 0 & 0
\end{array}}
\right) \ , \quad 
G_t'(k_1 = 1, k_2 = 1) = 
\left({\scriptsize
\begin{array}{lllll}
 1 & 0 & 0 & 0 & 1 \\
 -1 & 0 & 1 & 0 & 0 \\
 -1 & 1 & 0 & 0 & 0
\end{array}}
\right)
\end{equation}
We see that the two are not related by any $SL(3;\IC)$ transformations and are thus inequivalent toric varieties. We draw these diagrams explicitly in \fref{f:toricE=5}.

The other two models of the $E=5$ quiver theories can be similarly treated.
For Model (3), we have that
\begin{eqnarray}
\nn
&G_t^{(3)}& = \ker(\ker
{\scriptsize
\left(\begin{matrix}
1 & 1 & 1 \\ k_1 & k_2 & -k_1-k_2 
\end{matrix}\right)} 
\cdot 
{\scriptsize \left(
\begin{array}{lllll}
 -1 & -1 & 1 & 1 & 0 \\
 0 & 1 & -1 & 0 & 0 \\
 1 & 0 & 0 & -1 & 0
\end{array}
\right)})
=
\ker( \{ -k_2, \ -k_1 - k_2, \ k_1 + k_2, \ k_2, \ 0 \})
\\
&=&
\mbox{gen}
{\tiny
\left(
\begin{array}{lllll}
 0 & 0 & 0 & 0 & k_2 \\
 k_2 & 0 & 0 & k_2 & 0 \\
 k_1+k_2 & 0 & k_2 & 0 & 0 \\
 -k_1-k_2 & k_2 & 0 & 0 & 0
\end{array}
\right)} =
{\tiny
\left(
\begin{array}{lllll}
 0 & 0 & 0 & 0 & 1 \\
 1 & 0 & 0 & 1 & 0 \\
 k_1+k_2 & 0 & k_2 & 0 & 0 \\
 -k_1-k_2 & k_2 & 0 & 0 & 0
\end{array}
\right)}
 \ ; \ \
\gcd(k_1,k_2,k_1+k_2) = 1 \ .
\end{eqnarray}
This, as above, gives an infinite family, parametrised by choices of $k_1$ and $k_2$, of possibilities for the moduli space $\CM$ and hence inequivalent theories.

Finally, for Model (4), we have
\begin{eqnarray}
\nn
&G_t^{(4)} = \ker(\ker
{\scriptsize
\left(\begin{matrix}
1 & 1 & 1 \\ k_1 & k_2 & -k_1-k_2 
\end{matrix}\right) 
}
\cdot 
{\scriptsize \left(
\begin{array}{lllll}
 -1 & 0 & 1 & 0 & 0 \\
 0 & 1 & -1 & 0 & 0 \\
 1 & -1 & 0 & 0 & 0
\end{array}
\right)})
=
\ker( \{ -k_2, \ -k_1, \ k_1 + k_2, \ 0, \ 0 \})
\\
\label{Gt5-4}
=&
\mbox{gen}
{\tiny
\left(
\begin{array}{lllll}
 0 & 0 & 0 & 0 & k_2 \\
 0 & 0 & 0 & k_2 & 0 \\
 k_1+k_2 & 0 & k_2 & 0 & 0 \\
 -k_1 & k_2 & 0 & 0 & 0
\end{array}
\right)} 
=
{\tiny
\left(
\begin{array}{lllll}
 0 & 0 & 0 & 0 & 1 \\
 0 & 0 & 0 & 1 & 0 \\
 k_1+k_2 & 0 & k_2 & 0 & 0 \\
 -k_1 & k_2 & 0 & 0 & 0
\end{array}
\right)}
 \ ; \ \
\gcd(k_1,k_2,k_1+k_2) = 1 \ .
\end{eqnarray}

\paragraph{An Interesting Family: }
Examining Model (1) of \fref{f:E=5full} and Model (4) of \fref{f:E=6full}, it is clear that for the general case of $G$ nodes and $E=G+2$ fields, there will always be a cyclically-directed $G$-gon graph with 1 edge having an extra pair of arrows in opposite directions. Thus all $E$ fields are bi-fundamentals and let us, without loss of generality, place the extra pair of bi-fundamentals between nodes 1 and $G$, and let the direction of the arrows be $1 \to 2$, $2 \to 3$, \ldots, $(G-1) \to G$ and $G \to 1$. The quiver and adjacency matrices are
\begin{equation}
\begin{array}{lcl}
\begin{array}{l}\epsfxsize =6cm\epsfbox{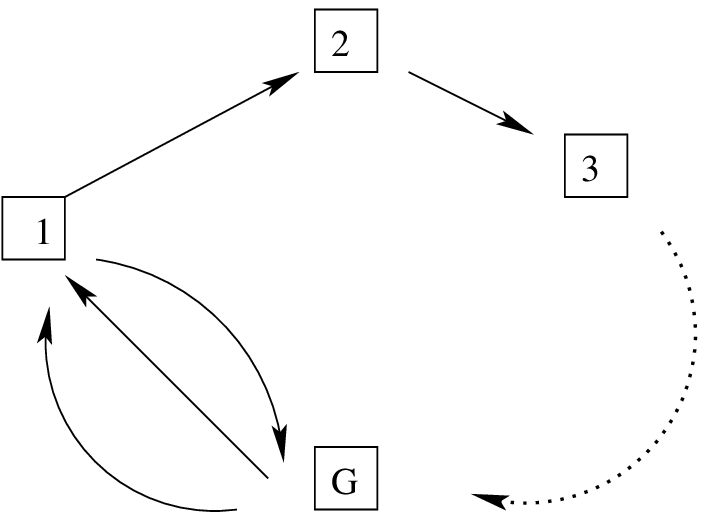}\end{array}
&~~&
d = 
\left(
\begin{matrix}
-1 & 0 & \ldots & 1 & 1 & -1 \\
1 & -1 & \ldots & 0 & 0 & 0 \\ 
0 & 1 & \ldots & 0 & 0 & 0 \\ 
\vdots & 0 & \vdots & \vdots & \vdots & \vdots \\
0 & 0 & \ldots & -1 & -1 & 1 \\
\end{matrix}
\right)_{G \times (G+2)}
\end{array}
\end{equation}
The 2-term superpotential is also straight-forward to write down:
\begin{equation}
W = \tr\left(
X_{12} X_{23} \ldots X_{G-2,G-1}
(X^1_{G-1,1} X_{1,G-1} X^2_{G-1,1} - X^2_{G-1,1} X_{1,G-1} X^1_{G-1,1})
\right) \ .
\end{equation}

We can also apply \eref{Gt} to the incidence matrix to find the toric diagram. The result has $E=G+2$ lattice points:
\begin{equation}
G_t  = \left(
\begin{matrix}
1 & 0 & -k_3 & - (k_3 + k_4) & \ldots & - (k_3 + \ldots + k_{G-1}) & 0 & 0 & -k_1 - k_2 \\
0 & 1 & k_2+k_3 & k_2+k_3+k_4 & \ldots &  k_2 + \ldots + k_{G-1} & 0 & 0 & k_1 \\
0 & 0 & 0 & 0 & \ldots & 0 & 1 & 0 & 1 \\
0 & 0 & 0 & 0 & \ldots & 0 & 0 & 1 & 1 \\
\end{matrix}
\right)_{4 \times (G+2)}
\end{equation}

\paragraph{Six Field Models: }
For completeness, let us present the matrices $G_t$ encoding the toric diagrams for the 6 models at $E=6$, drawn in \fref{f:E=6full}. They are, respectively,
\begin{equation}\label{Gt6}
\begin{array}{ll}
G_t^{(4)} = 
{\tiny
\left(
\begin{array}{llllll}
 1 & 0 & 0 & 0 & 0 & 1 \\
 k_1+k_2+k_3 & 0 & k_2 & 0 & k_2+k_3
   & 0 \\
 -k_1 & 0 & k_3 & k_2+k_3 & 0 & 0 \\
 -1 & 1 & 0 & 0 & 0 & 0
\end{array}
\right)
} \ , \quad
G_t^{(6)} = 
{\tiny
\left(
\begin{array}{llllll}
 k_1+k_2 & 0 & -k_3 & 0 & 0 & k_2 \\
 1 & 0 & 1 & 0 & 1 & 0 \\
 -k_1 & 0 & k_3 & k_2 & 0 & 0 \\
 -1 & 1 & 0 & 0 & 0 & 0
\end{array}
\right)} \\
\\
G_t^{(7)} = 
{\tiny
\left(
\begin{array}{llllll}
 0 & 1 & 0 & 0 & 0 & 1 \\
 1 & 0 & 0 & 0 & 1 & 0 \\
 k_1+k_2+k_3 & -k_3 & 0 & k_2 & 0 & 0 \\
 -k_1-k_2-k_3 & k_3 & k_2 & 0 & 0 & 0
\end{array}
\right)
} \ , \quad
G_t^{(10)} =
{\tiny
\left(
\begin{array}{llllll}
 1 & 0 & 0 & 0 & 0 & 1 \\
 k_1+k_2+k_3 & -k_3 & 0 & 0 & k_2 & 0 \\
 -k_1-k_3 & k_3 & 0 & k_2 & 0 & 0 \\
 -1 & 1 & 1 & 0 & 0 & 0
\end{array}
\right)}\\
\\
G_t^{(11)} = 
{\tiny
\left(
\begin{array}{llllll}
 0 & 0 & 0 & 0 & 0 & 1 \\
 1 & 0 & 0 & 0 & 1 & 0 \\
 1 & 0 & 1 & 1 & 0 & 0 \\
 -1 & 1 & 0 & 0 & 0 & 0
\end{array}
\right)
} \ , \quad
G_t^{(16)} = 
{\tiny
\left(
\begin{array}{llllll}
 0 & 0 & 0 & 0 & 0 & 1 \\
 0 & 0 & 0 & 0 & 1 & 0 \\
 k_1+k_2+k_3 & k_2 & 0 & k_2+k_3 & 0
   & 0 \\
 -k_1 & k_3 & k_2+k_3 & 0 & 0 & 0
\end{array}
\right)
}
\end{array}
\end{equation}

\subsection{A New Toric Duality}
In its original guise, toric duality \cite{Feng:2000mi} referred to the phenomenon of 4-dimensional, $\CN=1$ gauge theories having the same vacuum moduli space as toric varieties, some classes of these were later realised to be Seiberg dualities. In our host of examples above, we have again encountered this duality, now in a more general setting.

Take the two models of $E=4$, they, even though having quite different quivers, share the same infrared moduli space as $\IC^4$. Perhaps more dramatic is the following pair: take Model (4) of $E=5$ at $(k_1, k_2) = (1,1)$ and inspect \eref{Gt5-4}, then take Model (16) of $E=6$ at $(k_1, k_2, k_3) = (1,0,1)$ and inspect \eref{Gt6}, we see that they are
\begin{equation}
G_t^{E=5,~{\rm Model}(4)}=
{\tiny
\left(
\begin{array}{lllll}
 0 & 0 & 0 & 0 & 1 \\
 0 & 0 & 0 & 1 & 0 \\
 2 & 0 & 1 & 0 & 0 \\
-1 & 1 & 0 & 0 & 0
\end{array}
\right)
} \ , \quad
G_t^{E=6,~{\rm Model}(16)}=
{\tiny
\left(
\begin{array}{llllll}
 0 & 0 & 0 & 0 & 0 & 1 \\
 0 & 0 & 0 & 0 & 1 & 0 \\
 2 & 0 & 0 & 1 & 0 & 0 \\
-1 & 1 & 1 & 0 & 0 & 0
\end{array}
\right)
} \ .
\end{equation}
We see that if we removed a repeated column in the latter, which does not influence the toric description, the moduli spaces of the two theories, which have different number of gauge group factors, different matter content and interactions and different Chern-Simons levels, are identical as toric varieties. Clearly there are infinitely many such cases and it is interesting to study the systematics of this phenomenon.

\section{Tilings: Dimers versus Crystals}\label{s:dimer}\setall
It is now well-established that the most convenient and elegant way of encoding the $(3+1)$-dimensional quiver gauge theory of D3-branes probing toric Calabi-Yau threefold singularities is through the formalism of dimer models, or, equivalently, brane-tilings \cite{Hanany:2005ve,Franco:2005rj,Feng:2005gw}. The afore-mentioned ``toric condition''  \cite{Feng:2002zw} of the superpotential is naturally interpreted as the bi-partite (2-colour) property of the tiling while the Calabi-Yau condition of the threefold, which compels the toric diagram to be planar, gives the inherent structure of periodic tiling of the 2-dimensional plane. One question which has emerged is how this would generalise to toric varieties of higher dimension. Indeed, proposals have been made which suggest that a 3-dimensional analogue of the dimer, a so-named ``crystal model'', should encode the toric Calabi-Yau 4-fold case \cite{Imamura:2008qs,Lee:2007kv,Franco:2008um}. Is this so for the 4-folds we have encountered in our investigation?

Surprisingly, we find all our above CS theories to afford 2-dimensional tilings, rather than the naively expected crystals which tile 3-dimensions. Let us recall that for the $(3+1)$-dimensional gauge theories, an important relation exists \cite{Franco:2005rj}:
\begin{equation}\label{euler}
N_T - E + G = 0 \ ,
\end{equation}
where $N_T$ is the number of terms in the superpotential, while $E$ and $G$, as above, are the number of fields and gauge group factors. The recognition of \eref{euler} as the topological Euler equation for the simplex decomposition of a genus 1 Riemann surface $\Sigma$, with a number $N_T$ of vertices, a number $E$ of edges and a number $G$ of faces was key to the birth of the dimer model. Indeed, the graph dual of this simplex is a periodic version of writing the quiver diagram together with the superpotential, the periodicity further supporting the existence of $\Sigma$ as a torus. 

How does this crucial relation read for our $(2+1)$-dimensional CS theories? Here, since we are only considering 2-term superpotentials, $N_T = 2$. Moreover, recall that $E = G+2$ from \eref{symp}. Therefore $2-(G+2)-G=0$ and \eref{euler} is still satisfied! This is not what one would expect from a crystal model which is not a periodic tiling of the plane but which is perhaps at first expected for all toric 4-fold theories. In summary, we have that
\begin{quote}{\em
All quiver Chern-Simons theories corresponding to a M2-brane probing a toric Calabi-Yau 4-fold, such that the superpotential has two terms, admit a dimer model (2d-tiling) description.}
\end{quote}

\section{A General Counting}\label{s:count}\setall
We could have continued the process of Section \ref{s:tax} {\it ad infinitum}, listing more and more graphs and then for each, construct possible superpotentials with 2 terms or fix various values of Chern-Simons levels to obtain infinite families of toric moduli spaces. This, though explicit, is perhaps not so illustrative, let alone computationally prohibitive. It would, however, be most enlightening if we could count, say, the number of possible quivers for a given number of nodes. In this section, let us give a generating function to perform this count; we will find an elegant result very much in the spirit of the Plethystic Programme \cite{pleth}.

\subsection{A Systematic Enumeration}
\begin{figure}[h]\begin{center}
\epsfxsize = 3cm\epsfbox{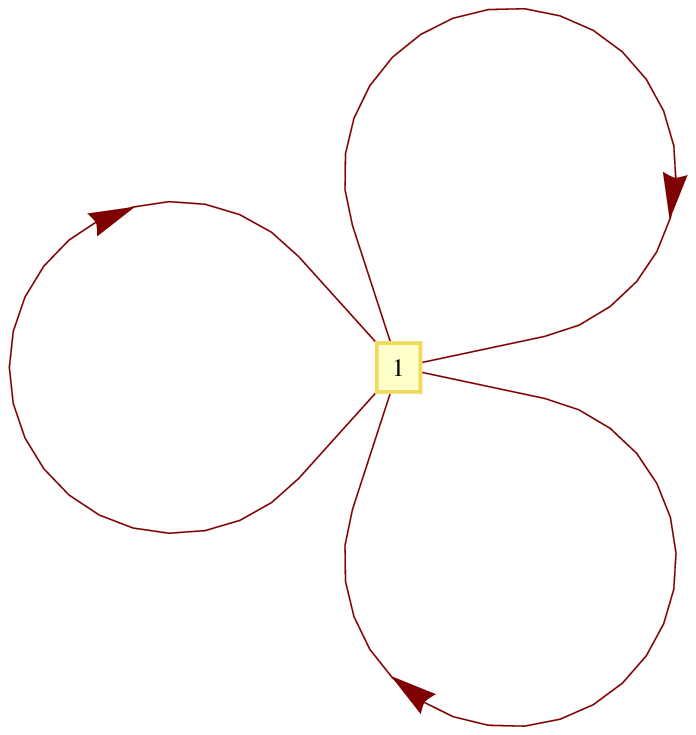}
\caption{{\sf The quiver for $\CN=4$ Super-Yang-Mills, corresponding to a D3-brane in flat $\IC^3$. This is a single-noded, triple-edged quiver, which we denote as 000.
}}
\label{f:C3}
\end{center}\end{figure}

To this end, we shall introduce a systematic enumeration and construction of the quivers. 
Let us do so by the concept of {\em base node} which we now introduce.
Examining Model (2) of \fref{f:E=4full}, Models (3) and (4) of \fref{f:E=5full}, as well as Models (7), (11) and (16) of \fref{f:E=6full}, we see that they, perhaps not immediately obviously, fall into a family. These are all models which have a single base node, viz., a single reference node whence loops depart and thence return.
Let us consider a chain of closed paths beginning and ending on this same node (corresponding to a possibly multi-trace gauge invariant operator) and denote it by a sequence of non-negative integers each entry of which encodes the length of one loop. Clearly, this sequence is unordered. For example, for the simple quiver which has a single node with 3 self-loops attached (incidentally, this is the quiver of the ${\CN=4}$ super-Yang-Mills theory for D3-branes in flat $\IC^3$ paradigmatic in the first AdS/CFT pair), we would denote it as 000, drawn in \fref{f:C3}.

Next, what would 001 (and of course any of its permutations) denote? This is Model (2) of \fref{f:E=4full}, which has 2 loops of length 0 and 1 loop of length 1. Note that by length here we mean the number of different nodes traversed before returning. Similarly, 011 and 002 correspond to Models (3) and (4) of \fref{f:E=5full} respectively. Likewise, we retrieve the 3 aforementioned models for $E=6$ fields corresponding to 111, 012 and 003. The systematics is thus clear. For $E$ fields and $G=E-2$ nodes, we are counting graphs which are in correspondence with unordered partitioning of $G-1$ into 3 parts of non-negative integers. This is a standard problem, whose solution is simply
\begin{equation}
f_1(t) = \frac{1}{(1-t)(1-t^2)(1-t^3)} = 1 + t + 2t^2 + 3t^3 + \ldots
\end{equation}
This function was indeed encountered in \cite{pleth} as the generating function for 1 adjoint field with $N=3$ D3 branes.
The above is the generating function such that the coefficient in front of $t^{G-1=E-3}$ gives the number of quivers of 1 base node with $G$ nodes. These are all quivers with 3 primitive loops.

Having addressed one family of quivers transcending across $E$, the family with a single base node, let us move onto 2 base nodes, say 1 and 2. Now, it is more convenient to count by open paths: we let $n_1n_2m_1m_2$ denote a configuration which has open paths of length $n_1$, $m_2$, $m_1$, $n_2$ respectively starting at node 1, ending at node 2, starting at node 2 and ending at node 1. Again, by length we mean distance away from the base pair. These are all quivers with 4 primitive paths between the 2 base nodes. Specifically, 0000 would correspond to Model (1) of \fref{f:E=4full}. Next, taking the base pair to be nodes 3 and 1, 0001 would denote Model (1) of \fref{f:E=5full}. Moving on to 4 nodes and 6 fields, we see that 0002, 0011 and 0101 denote respectively Models (4), (6) and (10) of \fref{f:E=6full} when taking nodes 4 and 1 and the base pair. This is analogous to the above, but is the unordered partitioning of $G-2$ into 4 non-negative integers with the extra complication that we must respect the di-hedral symmetry. Specifically, the transpositions $(n_1 \leftrightarrow n_2)$, $(m_1 \leftrightarrow m_2)$, $(n_1n_2 \leftrightarrow m_1m_2)$ generate a dihedral group of order 8. Orbits under these 8 elements must be quotiented out. The generating function for this also admits a standard solution by method of Molien series \cite{pleth} for the dihedral group, in a 4-dimensional representation acting on our 4-vector:
\begin{equation}
f_2(t) = \frac{1-t^6}{(1-t)(1-t^2)^2(1-t^3)(1-t^4)}
= 1 + t + 3t^2 + 4t^3 + 8t^4 + \ldots
\end{equation}
Here, the coefficient of $t^{G-2}$ is the number of models with $G = E-2$ nodes.

What about triplets or more of base nodes? Note that our above two have already exhausted all the quivers we have so far explicitly constructed. Could there be more families which one might encounter at higher $E$? We now argue that we shall, in fact, not. 
First, let us recall that our counting procedure above of course does not include any disconnected graphs and moreover excludes shapes such as Model (8) of \fref{f:E=6disconnect}. This is a 3-base-node example; however, we have explicitly shown that the superpotential vanished. Indeed, the cases of base nodes being one or two precisely permitted us to write a commutation relation allowing for the two terms in the superpotential: the case of the 1-base-node achieved with the adjoint and the case of the 2-base-node, with the pair of bi-fundamentals between them. Any other case must either be reducible to these two situations or have vanishing superpotential. 

Indeed, we can use an additional symmetry to restrict the possible models. Consider the number of flavours $n_E^i$ for a given node $i$: this is either an incoming-outgoing pair of arrows or an adjoint. 
The vector $n_E^i$ for $i=1,\ldots,G$ is an additional order parameter. From our examples, we see that it is 1 for most nodes. This is the one-flavoured node we discussed earlier. However, it must be at least 1 for every node since we are not counting disconnected quivers and it must never exceed 3 since there are only $E=G+2$ number of arrows in total. Clearly, the sum of $n_E^i$ over $i$ is equal to $E=G+2$. Therefore, there are only 2 possibilities for the components of the vector: (a) $G-1$ of them being 1 and a single 3 or (b) $G-2$ of them being 1 and 2 of them being 2.

Subsequently, this places a significant restriction on the total number of adjoints. Since each node must have at least one bi-fundamental, case (a) could have up to 2 adjoints on the last node and case (b) could have up to a single adjoint on the last two nodes. Hence, in total there could really only be 0, 1 or 2 adjoints. Now, we see that any node which has no adjoint and only an incoming/outgoing pair of arrows is a descendent of a simpler configuration: namely replace this node, together with its 2 attached arrows, with a single arrow between the 2 nodes from which the said node emanates. Therefore, we need only consider the parent quivers of (a) and (b), viz., a single node with flavour number 3 or two nodes each with flavour number 2, respectively. The former is 000. The latter has two possibilities: 0000 and a quiver which looks like $\epsfxsize =3cm\epsfbox{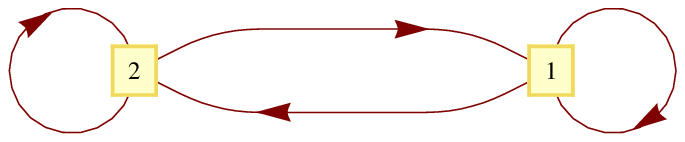}$; we can easily show that the latter admits no non-vanishing 2-term superpotential. Indeed, Model (8) of \fref{f:E=6disconnect} discussed above is a descendent thereof and also needs to be eliminated.

In summary, we only need to consider descendents of the 3-vector 000 and the 4-vector 0000 and no quivers with more than 2 base nodes survive.
Hence, $f_1$ and $f_2$ together exhaust all counting. Furthermore, noticing the shift in power of $f_2$ with respect to $f_1$, we conclude that:
\begin{quote}
{\em The number of theories of our interest, i.e., all non-trivial, connected models with 2 terms in a non-vanishing superpotential and $G = E-2$ nodes, is counted by the coefficient of $t^{G-1}$ in the expansion of the generating function}
\begin{equation}\label{gen}
f(t) = f_1(t) + tf_2(t) = \frac{1 + t + t^3 - t^4 + t^5}{(1-t)(1-t^2)(1-t^3)(1-t^4)} \ .
\end{equation}
\end{quote}

\subsection{A Geometric Aper\c{c}u}
The forms of $f_1$ and $f_2$ are perhaps familiar to the astute reader. The first, is the Hilbert series (cf.~\cite{pleth}) for the weighted projective plane $W\IP^2_{[1:2:3]}$ (or, alternatively the quotient $\IC^3/S_3$ of the symmetric group on 3 objects), and the second, that of the 4-fold quotient singularity $\IC^4 / D_4$, i.e., the orbifold of $\IC^4$ by the dihedral group of order 8 (cf. \cite{Hanany:1999sp} for discrete subgroups of $SU(4)$). It is quite elegant that these two geometrical spaces should encode the entire space of quivers of our type.

In fact, quivers of the first family, the single-base-node type, obey a simple rule of composition, reflecting the fact that there is an underlying commutative algebra which is freely generated: the zero element is the self-adjoining loop, 1 is the quiver with 2 nodes and a pair of bi-fundamentals in opposite directions between them, and thus generalising to $n$, which is the quiver that is an $n$-gon with bi-fundamentals cyclically going around once. We can create a quiver by selecting any 3 from this list and composition is by pasting the three at a chosen common node:
\[
\epsfxsize = 13cm\epsfbox{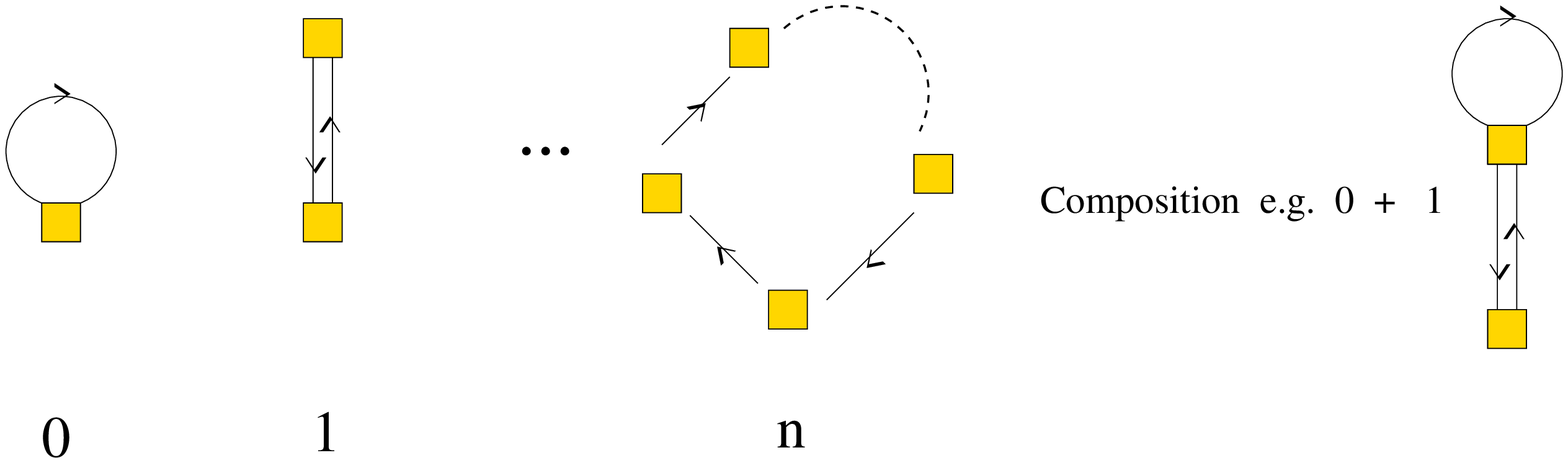}
\]
In terms of the 3-vector of integers, we can also see this composition. For example, 001 + 011 = 012 refers to the following pasting of quivers
\[
\epsfxsize = 12cm\epsfbox{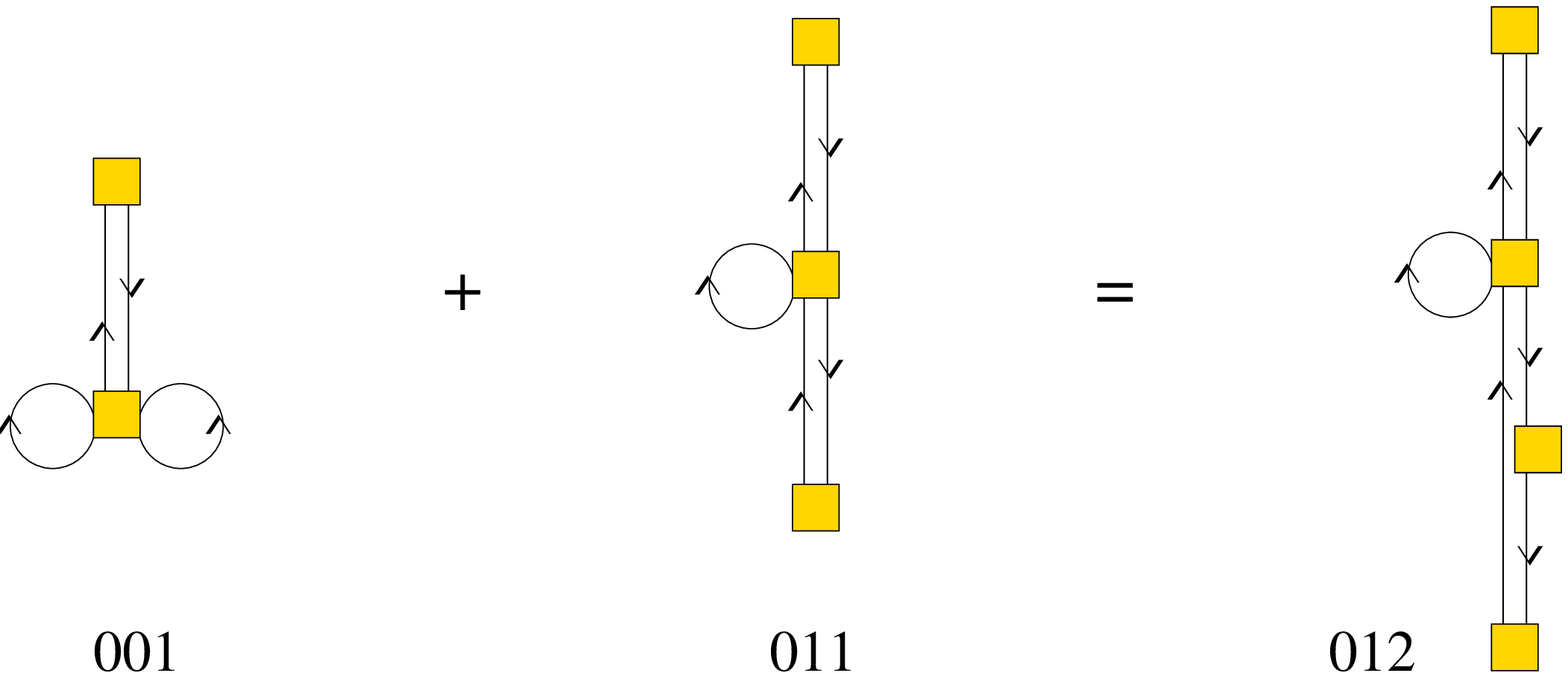}
\]
Indeed, we have drawn the result in a suggestive form, so as to indicate that such a composition corresponds to the insertion of nodes. Specifically, adding $n_1n_2n_3$ signifies inserting, after choosing an absolute ordering of the loops, $n_1$, $n_2$ and $n_3$ nodes into the loops. In this fashion, we readily see that the quivers $001$, $011$ and $111$ generate, by this insertion procedure, all quivers with one base node. Thus, we have an algebra freely generated by the three elements:
\[
\epsfxsize = 14cm\epsfbox{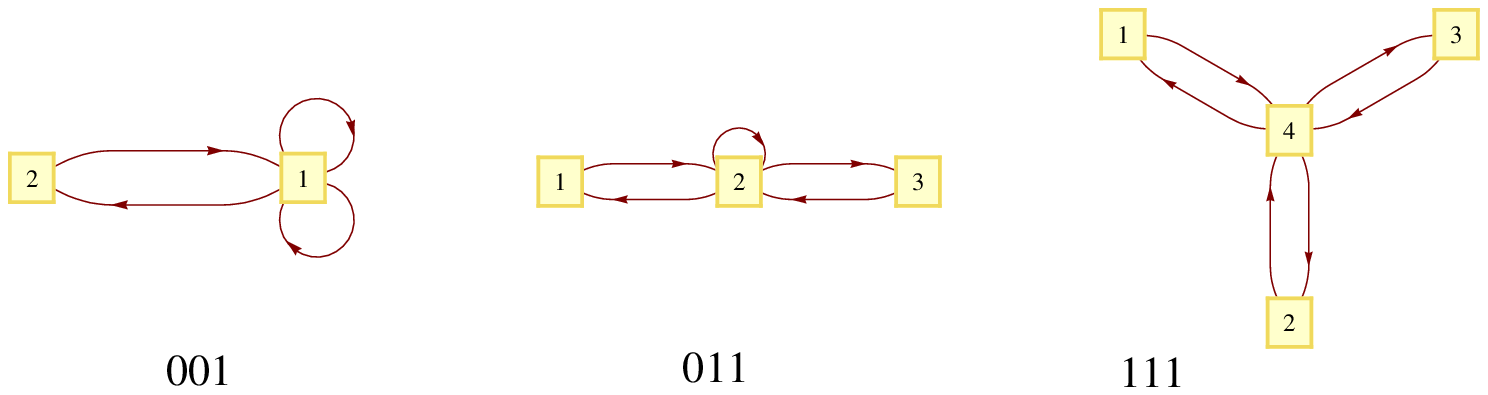}
\]

The second family is a little more involved. We can define the 0 element as a node with an out-going arrow, 1 as a node with an incoming and an out-going arrow, etc.:
\[
\epsfxsize = 11cm\epsfbox{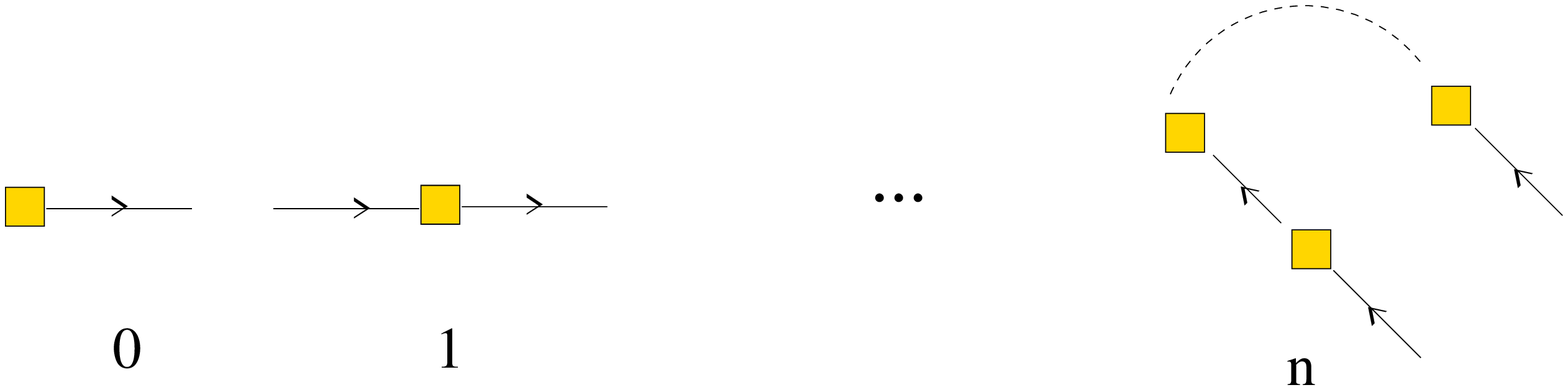}
\]
Composition is by again pasting, though this time we must first paste to create a pair of base nodes, unto which we may then attach the open arrows of the element $n$. Finally, we must eliminate the redundancies of quivers which obey dihedral symmetry. In terms of the 4-vector, we now have an algebra generated by 
5 elements, 0001, 0011, 0101, 0111 and 1111:
\[
\epsfxsize = 16cm\epsfbox{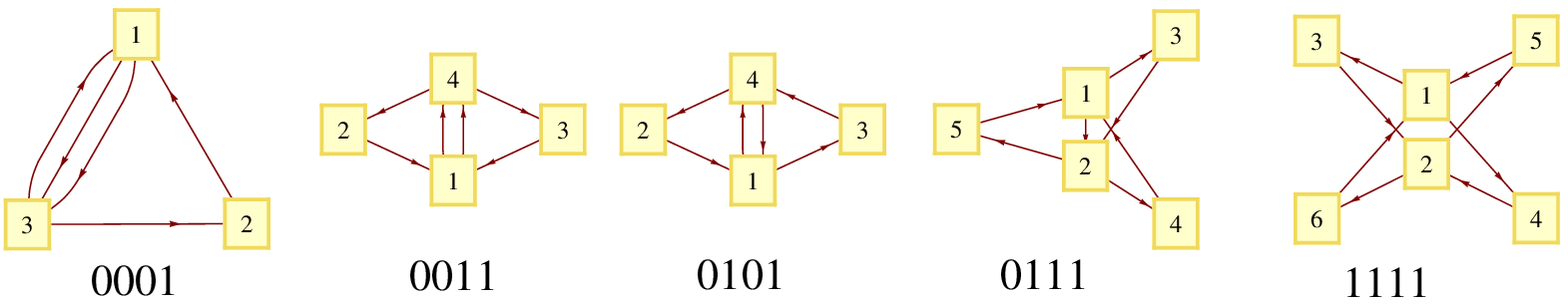}
\]
obeying a single relation.

We have thus uncovered an algebraic structure on the space of CS quivers with 2-term superpotentials. This is expected to persist to higher number of terms and indeed to the space of all quivers; we have thus found an underlying algebraic variety for the set of Chern-Simons quivers.
We remark also that taking the plethystic logarithm \cite{pleth} of the total $f(t)$ reveals that it is a non-terminating series suggesting that the algebraic variety is not a complete intersection. It is in the form of a set union of the two orbifolds, though there is a degree shift on the second.

\section{Conclusions and Prospects}\label{s:conc}\setall
In this paper we have started a taxonomic study of quiver Chern-Simons theories in $(2+1)$-dimensions. In particular, we study the theories arising from M2-branes probing toric Calabi-Yau 4-fold singularities. Our purpose is not to merely provide a catalogue of such theories since there are clearly infinite families thereof, but to study the space of these theories from both a synthetic and an analytic approach, and to uncover interesting physical phenomena as well as mathematical structure. Some of the techniques and concepts can indeed be generalised to quiver theories in arbitrary dimension and with any supersymmetry.

We have first presented, in flow-chart \eref{Gt}, a succinct way of a ``forward algorithm'' which takes the quiver and superpotential data as input and the moduli space as an output: the former as a pair of matrices, respectively the incidence and the perfect matching matrices, and the latter, as a matrix representing the integer coordinates of the toric diagram of the Calabi-Yau 4-fold. The intermediate computations involve nothing more than matrix multiplications and finding kernels. Indeed, this generalises and significantly improves on the forward-algorithm of the case of $(3+1)$-dimensional gauge theories of \cite{Feng:2000mi}: we only need to insert a $C$-matrix of the Chern-Simons levels and have obviated the need for finding dual cones.

Thus armed, we have commenced the classification of all quiver Chern-Simons theories. Listing the quivers is matter of combinatorics and for simplicity we have taken the superpotential to have two terms. It is clearly a pressing problem what happens at an arbitrary number $N_T$ of terms, this we shall leave to forth-coming work. Subsequently our order parametre becomes $G$, the number of nodes and $E$, the number of arrows, is equal to $G+2$ since the master space here is $\IC^E$. In general, we should have 3 order parametres $(E,G, N_T)$. We exhaustively list the first members at $E=4,5,6$ and find many non-trivial theories, including and extending beyond what has so far emerged in the literature.

Because of our forward algorithm, we can readily determine the toric Calabi-Yau 4-fold moduli spaces for all the theories presented. These exhibit as many toric varieties, each of which is an infinite family, indexed by the integer Chern-Simons levels. Remarkably we have once more encountered the phenomenon of ``toric duality'' first noted in \cite{Feng:2000mi} for $(3+1)$-dimensional gauge theories. Now, it has manifested in perhaps some more striking guises: we find not only theories with the same number of nodes but also those with different number of nodes, fields, as well as Chern-Simons levels, to flow to the same IR moduli space as toric Calabi-Yau 4-folds. It is certainly of importance to study what is precisely happening in the field theory of such pairs.

Building upon our host of examples, we have stepped back for a panoramic view of all our theories: connected quiver Chern-Simons with 2-term toric superpotentials. Though the graphs seem unwieldy and growth rapidly in number as we increase the number of nodes, we have found a generating function \eref{gen} which counts the number of inequivalent theories given the number of nodes. Remarkably, this generating function splits conveniently into 2 pieces, each of which is the Hilbert series of a precise algebraic variety: the non-Abelian quotient space $\IC^3/S_3$ by the symmetric group on 3 objects of order 6 and the Abelian quotient $\IC^4 / D_4$  by the ordinary dihedral group of order 8 (i.e., the symmetry group of the square). Could we package the case with arbitrary $(E,G,N_T)$ into a convenient tri-variate generating function and would the result have an interpretation as a Hilbert Series, whereby suggesting that the space of quiver theories is itself some algebraic variety? In answering this and many questions raised we are presently engaged. Indeed, we have only tread upon the fringe of a vast and fertile land, into her bountiful bosom we must onwardly march.

\section*{Acknowledgments}
{\it Scientiae et Technologiae Concilio Y.-H.~H. hoc opusculum dedicat cum gratia ob honorem officiumque Socii Progressi collatum. Et Ricardo Fitzjames, Episcopo Londiniensis, ceterisque omnibus benefactoribus Collegii Mertonensis Oxoniensis quorum beneficiis pie, studiose, iucunde vivere licet, pro amore Catharinae Sanctae Alexandriae et ad Maiorem Dei Gloriam.}

\appendix
\section{Preliminary Classification}
\begin{figure}[ht]\begin{center}
\epsfxsize = 10cm\epsfbox{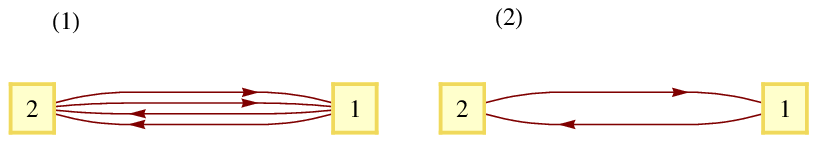} 
\caption{{\sf The quivers with 4 fields and 2 nodes. There are 2 solutions.
Note that Model (2) has only 2 arrows because it has 2 adjoint fields which are not drawn.
}}
\label{f:E=4}
\end{center}\end{figure}

\begin{figure}[ht]\begin{center}
\epsfxsize = 14cm\epsfbox{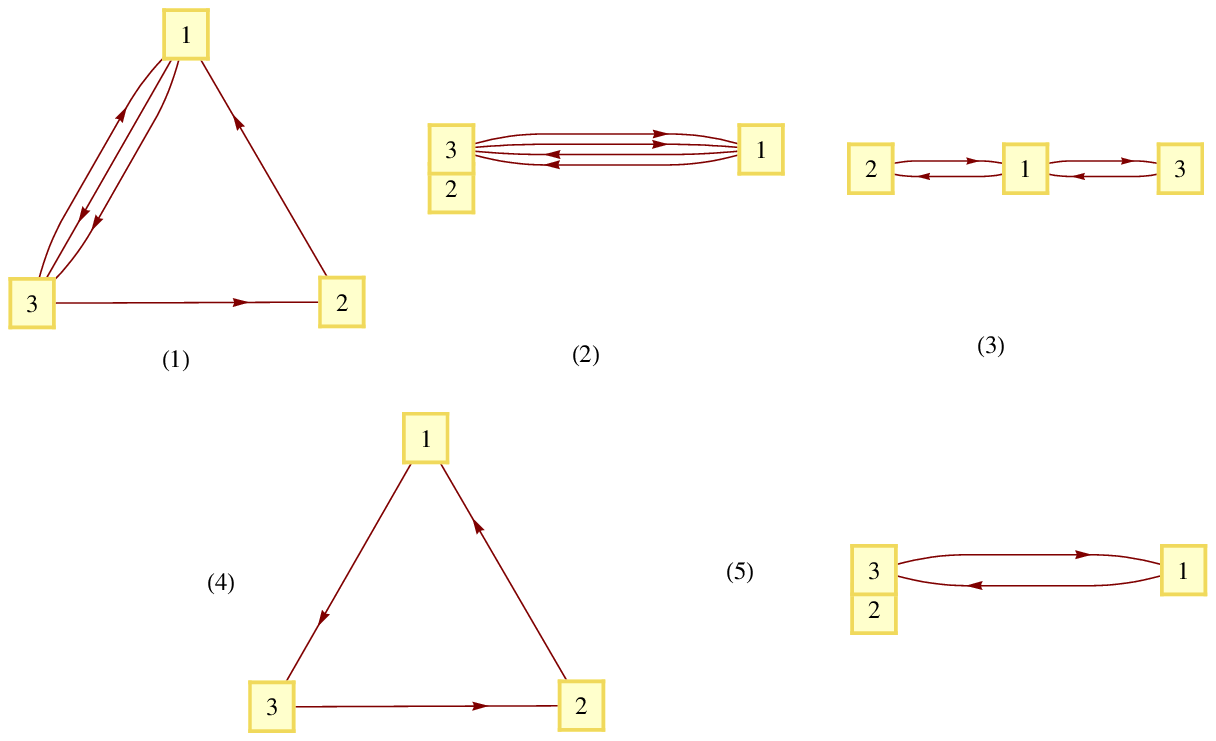} 
\caption{{\sf The quivers with 5 fields and 3 nodes. There are 5 solutions. Note that all quivers with explicitly less than 5 arrows have adjoint fields which are not drawn.}}
\label{f:E=5}
\end{center}\end{figure}

\begin{figure}[ht!!!]\begin{center}
\epsfxsize = 16cm\epsfbox{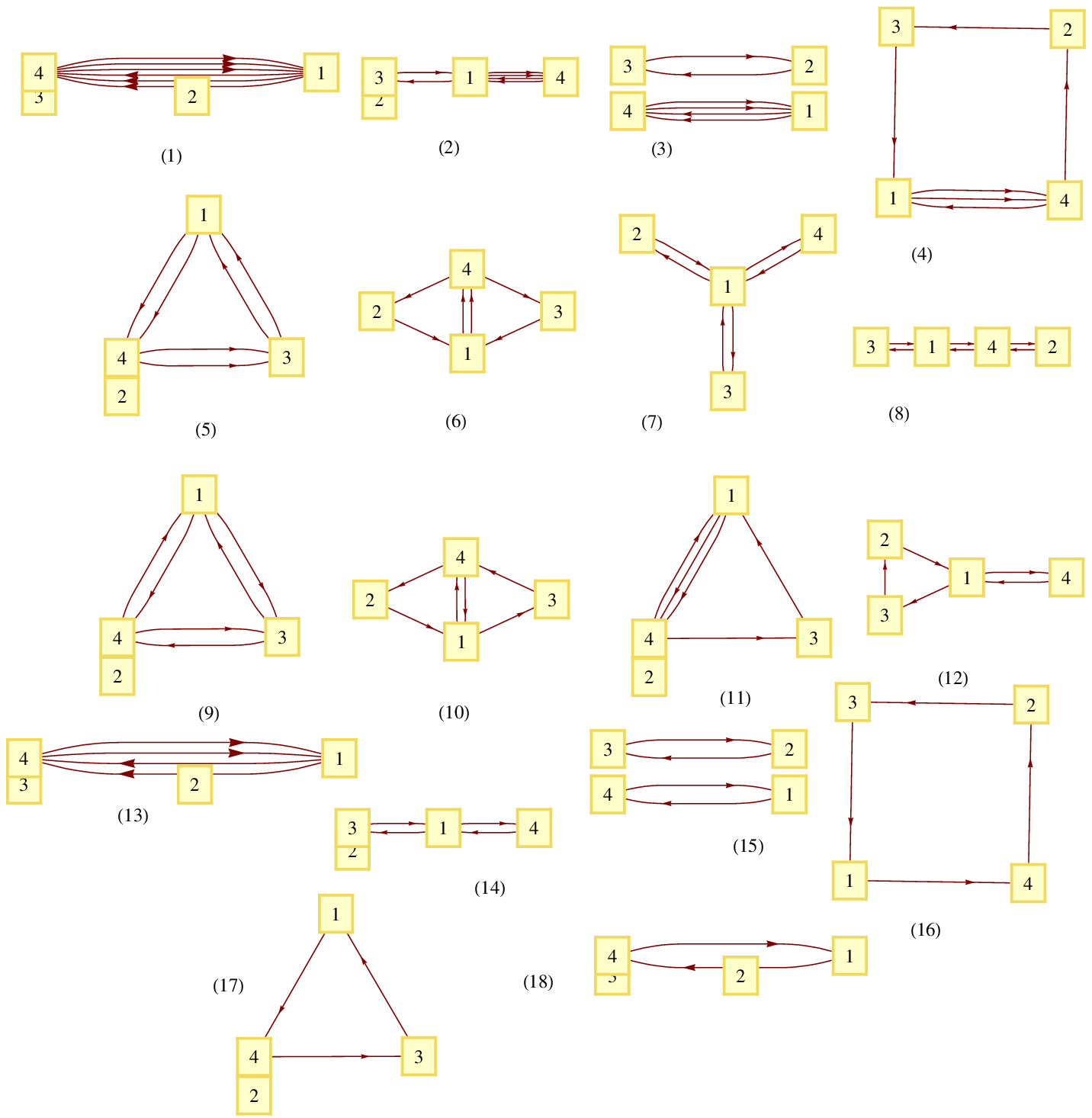} 
\caption{{\sf The quivers with 6 fields and 4 nodes. There are 18 solutions. Note that all quivers with explicitly less than 6 arrows have adjoint fields which are not drawn.}}
\label{f:E=6}
\end{center}\end{figure}

\setall
In this appendix, we present the results for a preliminary classification of the case of 2 terms in the superpotential. This will include cases of disjoint quivers and unmarked adjoint fields, we present them here for completeness. The cases of 4, 5 and 6 fields are respectively shown in Figures \ref{f:E=4}, \ref{f:E=5} and \ref{f:E=6}. In the text we will sift through these graphs carefully, adding adjoints wherever consistent.

\section{Perfect Matchings from the Superpotential}\label{a:P}
In this appendix, we extract the matrix $P$ of perfect matchings directly from the superpotential $W(\{X_i\})$ of the fields $X_{i=1,\ldots,E}$. In the archaic days of the forward algorithm, $P$ is found to be the product of the matrix $K$ (which encodes how a total of $E$ F-terms can be solved in terms of a fewer number) with its dual cone matrix $T$. Now we can follow the ensuing prescription to directly obtain $P$ without the rather computationally intensive procedure of finding dual cones:
\begin{enumerate}
\item
Group all the monomials appearing in $W$ into those with positive coefficients and those with negative, giving us two sets of equal length (since each field must appear exactly twice with opposite signs by the toric condition), on each of which we choose an order;

\item
Find the position of field $X$ in each of the two sets; say it is $i$-th in the positive set and $j$-th in the negative set. Construct a matrix whose $(i,j)$-th entry is $X$ (should the index ever repeat, simply add, to the entry, the new field). Do so for all $E$ fields and patch in zeros otherwise. This is the Kasteleyn matrix. Compute its determinant $D$.

\item 
$D$ is a new polynomial in the $E$ fields with $c$ (the number of perfect matchings) terms. Fix an order for these terms, and construct and $E \times c$ matrix. For each of the $E$ fields find the location in the monomial in $D$. We place a 1 at this location in the $E$-th row correspondingly and 0 otherwise. The resulting matrix is the desired $P$.

\end{enumerate}

As an illustrative example, let is consider the famous cone over the zeroth del Pezzo surface, i.e., the Abelian orbifold $\IC^3 / \IZ_3$. Its quiver is a directed loop over 3 nodes, each with multiplicity 3 and its superpotential is a six term cubic in 9 fields:
\begin{equation}
W = \sum\limits_{i,j,1}^3 \epsilon_{ijk} X_{1,2}^iX_{2,3}^jX_{3,1}^k \ .
\end{equation}
We recall that (cf.~e.g., Eq. (2.12) - (2.17) of \cite{master}) the explicit solutions to the F-terms are ${{X_{1,2}^1}= 
    {\frac{X_{1,2}^3\,X_{3,1}^1}{X_{3,1}^3}}}, \
  {{X_{1,2}^2}= 
    {\frac{X_{1,2}^3\,X_{3,1}^2}{X_{3,1}^3}}}, \ 
  {{X_{2,3}^1}= 
    {\frac{X_{2,3}^3\,X_{3,1}^1}{X_{3,1}^3}}}, \
  {{X_{2,3}^2}= 
    {\frac{X_{2,3}^3\,X_{3,1}^2}{X_{3,1}^3}}}$ and hence 
\begin{equation}
K = 
{\scriptsize
\ba{c|ccccccccc}
& X_{1, 2}^1 & X_{1, 2}^2& X_{1, 2}^3& X_{2, 3}^1& X_{2, 3}^2& X_{2,
  3}^3&  X_{3, 1}^1& X_{3, 1}^2& X_{3, 1}^3 
\\ \hline
X_{1, 2}^3 & 1 & 1 & 1 & 0 & 0 & 0 & 0 & 0 & 0 \\
X_{2, 3}^3 & 0 & 0 & 0 & 1 & 1 & 1 & 0 & 0 & 0 \\
X_{3, 1}^1 & 1 & 0 & 0 & 1 & 0 & 0 & 1 & 0 & 0 \\
X_{3, 1}^2 & 0 & 1 & 0 & 0 & 1 & 0 & 0 & 1 & 0 \\
X_{3, 1}^3 & -1 & -1 & 0 & -1 & -1 & 0 & 0 & 0 & 1
\ea} \ ,
\quad
T = \mbox{Dual}(K) = 
{\scriptsize
\left( \begin{matrix}
   0 & 0 & 1 & 1 & 0 & 0 \cr
   0 & 0 & 1 & 0 & 1 & 0 \cr
   1 & 0 & 0 & 0 & 0 & 1 \cr
   0 & 1 & 0 & 0 & 0 & 1 \cr
   0 & 0 & 1 & 0 & 0 & 1 \cr  \end{matrix} \right) \ ,
}
\end{equation}
where Dual refers to finding the dual cone. The product $K^t \cdot T$ is the matrix $P$.

Using our new algorithm, we find that there are 3 positive terms and 3 negative terms, giving us a Kasteleyn matrix and its determinant $D$ as:
\begin{equation}
Kas = {\scriptsize \left(
\begin{array}{lll}
 X_{3,1}^1 & X_{2,3}^3 & X_{1,2}^2 \\
 X_{1,2}^3 & X_{3,1}^2 & X_{2,3}^1 \\
 X_{2,3}^2 & X_{1,2}^1 & X_{3,1}^3
\end{array}
\right)} \ , \qquad
\begin{array}{rcl}
D &=& X_{1,2}^1 X_{1,2}^2 X_{1,2}^3-X_{2,3}^3 X_{3,1}^3 X_{1,2}^3+X_{2,3}^1 X_{2,3}^2 X_{2,3}^3\\
&&-a(1) X_{2,3}^1 X_{3,1}^1-X_{1,2}^2 X_{2,3}^2 X_{3,1}^2+X_{3,1}^1 X_{3,1}^2 X_{3,1}^3
\end{array}
\end{equation}
Upon comparison, we find that both procedures give (up to trivial permutation of columns, note that we have fixed the row by a canonical ordering of the fields) $P =$ {\tiny $\left(
\begin{array}{llllll}
 1 & 0 & 1 & 0 & 0 & 0 \\
 1 & 0 & 0 & 1 & 0 & 0 \\
 1 & 0 & 0 & 0 & 1 & 0 \\
 0 & 1 & 1 & 0 & 0 & 0 \\
 0 & 1 & 0 & 1 & 0 & 0 \\
 0 & 1 & 0 & 0 & 1 & 0 \\
 0 & 0 & 1 & 0 & 0 & 1 \\
 0 & 0 & 0 & 1 & 0 & 1 \\
 0 & 0 & 0 & 0 & 1 & 1
\end{array}
\right)$}.

We must point out that on a simple timing contrast, our new algorithm for this simple example is about a factor of 10 faster! This is significant since the algorithm for finding dual cones is exponential running time so for bigger examples, we expect that our new method to be a substantial improvement. For example, for the cone over the third del Pezzo surface, with 14 fields and 8 terms in the superpotential, our present algorithm constitutes a time reduction of more than 1000-fold over the old dual-cone method! It would be interesting to see whether this technique could be generalised to rapidly find the dual cone of arbitrary integer cones; this would of tremendous use to toric geometry.

%
%

\end{document}